\documentclass[fleqn,usenatbib]{rasti}

\usepackage{newtxtext,newtxmath}

\usepackage[T1]{fontenc}

\DeclareRobustCommand{\VAN}[3]{#2}
\let\VANthebibliography\thebibliography
\def\thebibliography{\DeclareRobustCommand{\VAN}[3]{##3}\VANthebibliography}

\usepackage{graphicx}	
\usepackage{amsmath}	
\usepackage{booktabs}

\newcommand{\changes}[1]{#1}

\title[Emulation of non-linear 1D spectral models]{Emulation of non-linear 1D spectral models: relativistic X-ray reflection}

\author[B. J. Ricketts, T. Hadži Veljković et al.]{
Benjamin J. Ricketts$^{1,2}$\thanks{E-mail: b.ricketts@sron.nl},
Tin Hadži Veljković$^{3}$ \thanks{Co-first author},
Daniela Huppenkothen$^{1}$,
Adam Ingram$^{4}$,
\newauthor{Matteo Lucchini$^{1}$, Guglielmo Mastroserio$^{5}$ and Fergus J. E. Baker$^{4}$} \\
$^{1}$Anton Pannekoek Institute, University of Amsterdam, Science Park 904, Amsterdam, 1098 XH, Netherlands\\
$^{2}$SRON, Niel Bohrweg 4, Leiden, 2033 CA, Netherlands\\
$^{3}$Informatics Institute, University of Amsterdam, Science Park 900, Amsterdam, 1098 XH, Netherlands\\
$^{4}$ Centre for Extragalactic Astronomy, Department of Physics, Durham University, South Road, Durham DH1 3LE, UK\\
$^{5}$ Scuola Universitaria Superiore IUSS Pavia, Palazzo del Broletto, piazza della Vittoria 15, I-27100 Pavia, Italy\\
}

\date{Accepted XXX. Received YYY; in original form ZZZ}

\pubyear{\the\year{}}

\begin{document}
\label{firstpage}
\pagerange{\pageref{firstpage}--\pageref{lastpage}}
\maketitle

\begin{abstract}
The use of machine learning techniques to approximate computationally expensive models has become increasingly prevalent in a wide variety of fields within astronomy. We discuss the implementation of emulators for 1-dimensional models in the context of the astrophysical numerical model \texttt{reltrans}, a black hole X-ray spectral model that models the effects of relativistically smeared emission from an accretion disk. We argue that the decision of whether and how to emulate should follow from a systematic characterisation of the target model, and we demonstrate a diagnostic workflow: examining how the spectrum varies with individual parameters. 
We adopt a modular strategy, emulating only the relativistically convolved reflection spectrum ($\approx1-10\%$ of the total flux) rather than the full model. Using an operator-learning architecture with Fourier feature embeddings and FiLM conditioning, we reproduce the reflection spectrum to $O(0.1)\%$ precision across 0.1-100 keV with a 4-10× speed-up that scales considerably better under vectorised evaluation. This emulator, \texttt{RTFAST2}, recovers the true parameters of simulated observations without the systematic posterior biases of our previous work. We conclude that no architecture is universally transferable and bespoke emulators motivated by a model's specific structure are required. The modular approach taken in this work presents a promising strategy for future emulators of numerical models.
\end{abstract}




\section{Introduction}

As our understanding of the physics in astronomical systems improves and we begin to use different types of observations (such as spectral variability and polarisation), our numerical models must begin to represent these systems more comprehensively. However, with additional physics in our models, the computational complexity and evaluation time generally increased, and it has become more difficult to reliably constrain correlations between the physical parameters added.

To explore the posterior probability of model parameters with respect to the data, we often use techniques such as Markov Chain Monte Carlo \citep{metropolis1953equation,10.1093/biomet/57.1.97} or nested sampling \changes{\citep{skilling2006,speagle2020dynesty,buchner2023nested}}. These techniques require on the order of hundreds of thousands to millions of model evaluations, and the number of evaluation scales strongly with the number of free parameters. While a simpler model that can be evaluated quickly might take a few minutes at most to sample the posterior, computationally expensive numerical models begin to take hours to days (in the extreme, months to years) to sample the posterior of a single dataset. Such slow analysis means that theoretical development is slower to be informed by new data and that models which attempt to better physically explain observed data are disfavoured by observers.

One possible solution to this problem is to emulate or approximate these complex models such that they become considerably cheaper to compute (with the trade-off of imprecise approximation of the true physical model). Historically, this was sometimes achieved by analytical functions (such as \texttt{nthcomp} \citep{zdziarski1996broad} ), but more recently, astronomy has begun to adopt neural networks and other machine learning techniques instead as they can better represent complex non-linear functions that emerge from physical numerical models than simple analytical functions. Examples include cosmological simulations \citep{heitmann2009coyote,bird2019emulator,kaushal2022necola}, cosmological matter power spectra \citep{heitmann2013coyote,spurio2022cosmopower}, optical and UV energy spectra of galaxies and Type Ia supernovae \citep{alsing2020speculator,kerzendorf2021dalek} and X-ray spectral studies of accretion disk winds in Active Galactic Nuclei (AGN) \citep{matzeu2022new}. Emulators have been built with a wide range of techniques, including: Gaussian processes \citep{bastos2009diagnostics}, artificial neural networks (ANNs) \citep{himes2022accurate}, and kernel ridge regression \citep{vicent2018emulation}. 

\begin{table*}
\centering
\begin{tabular}{lllll}
Parameter & Description &Space & Range & Units                  \\
\hline
h         & lampost corona height &log     &2 - 700 & $R_g$                     \\
a         & spin &linear  &0 - 0.998 & *                      \\
Inclination       & inclination &log &  1 - 89   & degrees                \\
$R_{\text{inner}}$    & inner disk radius &log    & 1 - 400 & ISCOs                  \\
$R_{\text{outer}}$    & radius of the outer edge of the disk &log    & 400 - $10^5$ & $R_g$                     \\
$\Gamma$     & photon index &linear & 1.4 - 3.4 & *                      \\
$\text{log}\xi$ & ionisation of the disk & linear & 0-4.7 & * \\
$A_{\text{Fe}}$       & iron abundance of disk  in solar units &log    & 0.5 - 10 & *                      \\
$\text{log}N_e$     & electron density of disk &linear  & 15 - 20 & $N_e$: cm$^{-3}$                      \\
kTe       & source electron temperature &log     & 30 - 500& keV               
\end{tabular}
\caption{List of parameters in \texttt{RTFAST2}, the type of parameter space used in training and their units. All units indicated with a * are dimensionless.}
\label{table:pars}
\end{table*}

\begin{table*}
    \centering
    \begin{tabular}{c c c c}
        Parameter & Description & Value & Note\\
        \hline
        $z$ & Redshift & 0 & Simple spectrum shift\\
        $n_H$ & Galactic absorption & 0 & Galactic absorption models are computationally cheap\\
        boost & Normalising factor for the reflection spectrum & -1 & -1 returns only the reflection spectrum in reltransDCp\\
        $M$ & black hole mass & $3\times10^6$ & Mass has no effect on the time averaged spectrum\\
        $norm$ & Overall normalisation & 1 & Total model multiplicative factor 
    \end{tabular}
    \caption{List of fixed parameters relevant to time-averaged spectra used in generation of data, the effects of which are instead analytically applied in total model calculation.}
    \label{table:fixed}
\end{table*}

Each use-case of emulation has different constraints. For instance, in emulations of cosmological simulations, the emulators can often still be computationally expensive (on order of minutes to compute) as an individual evaluation of the ground truth model can take as long as a month to compute. Data are often scarce in these scenarios, so larger, more complex strategies are favoured in order to best make use of the few evaluations that can be used as training data. This is very different from use-cases where numerical models can evaluate much faster ($\approx1$ second) but where either the model is to be used to sample posterior probabilities for many different data sets, or posteriors are complex enough to require potentially millions of model evaluations, and where speed-ups of a factor of ~10 represent a substantial improvement. 

Presently, we focus on the second case, and in particular to the model \texttt{reltrans} \citep{ingram2019public,mastroserio2021modelling,ingram2022,2023ApJ...951...19L}, an X-ray reverberation and spectroscopic model of X-ray reflection around black holes (see also Section \ref{sec:reltrans} for more details). In previous work (\citet{2025MNRAS.538.1096R}, subsequently BR25), we trained an emulator (\texttt{RTFAST}) for the \texttt{RTDIST} flavour\footnote{A version of reltrans which allows for constraining of distance to the observed source utilizing the self-consistent calculation of the ionisation of an accretion disk surrounding the object} of \texttt{reltrans} to an average precision of $3\%$ with a minimum $\mathcal{O}(10^2)$ speed up over the original numerical model over all 17 physically meaningful parameters in the model. Given that instrumental systematics in X-rays are of the same order of magnitude as the emulator error, the latter could in general be considered satisfactory. However, we found that the emulator would retrieve very confident, but biased posteriors that were off by a few percent of the true parameter in simulated observations. While this would not change the physical interpretation in a particularly meaningful manner, it is nevertheless important to reduce the effect of emulator bias to a minimum. In addition, \texttt{RTFAST} was only usable with XMM-Newton observations due to being trained on a fixed energy grid between 0.1 and 20 keV.

In this paper, we present an update to \texttt{RTFAST}: in \texttt{RTFAST2}, we emulate not the full model, but only the computationally expensive reflection spectrum (Section \ref{sec:reltrans}). We also incorporate recent innovations from the machine learning literature into our architecture (Section \ref{sec:emulator}), leading to a factor of ~10 increase in accuracy, though at the trade-off of a factor of ~10 decrease in computational performance compared to \texttt{RTFAST}. These innovations also allow for the generalisation of the emulator to all relevant instruments with arbitrary energy resolution.

We will also discuss the various trade-offs and considerations necessary when building emulators for complex astrophysical models, and show throughout the paper that thoughtful consideration of the physical model and its properties leads to more robust and trustworthy emulator models.

\section{Reltrans and Training Data Generation} \label{sec:reltrans}

\begin{figure*}
    \centering
    \includegraphics[width=0.95\linewidth]{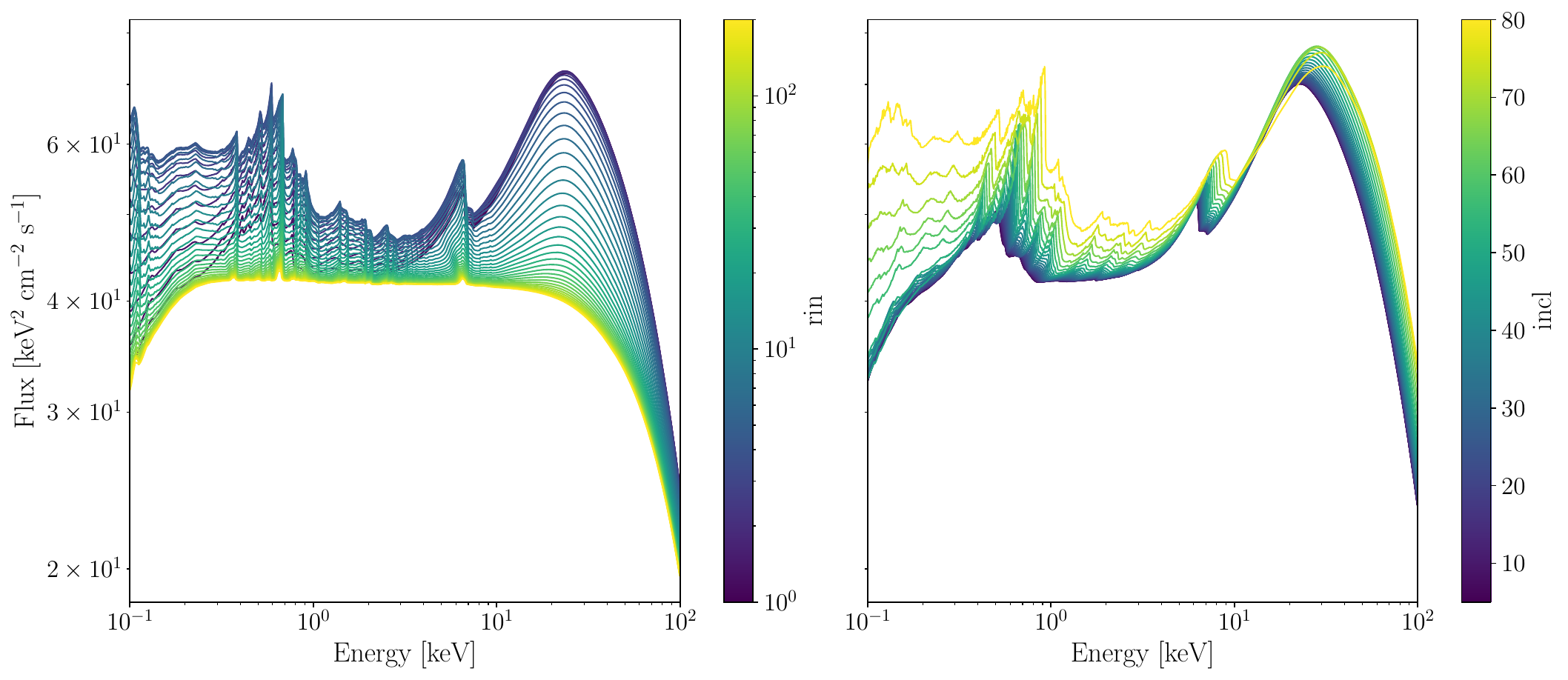}
    \caption{Left: the reflection spectrum coloured as a function of how far the inner-most radius of the disk extends. Dark blue indicates smaller truncation of the disk and more reflection and relativistic effects, bright yellow indicates larger truncation and less reflection and relativistic effects. Right: the reflection spectrum coloured as a function of inclination of the disk in respect to the observer. The larger the inclination, the closer to edge on the disk is.}
    \label{fig:spectra_grid}
\end{figure*}

Here, we will briefly discuss \texttt{reltrans} and refer the reader to \citet{mastroserio2021modelling} for a more comprehensive explanation. This model is non-linear in its response to the parameters, and thus serves as a good example of how non-linear models with high- and low-frequency features in the spectrum across a large parameter space can be approached with emulators. In practical terms, using \texttt{reltrans} in inference is computationally expensive (around 0.4 second per model evaluation for the time-averaged spectrum) due to on-the-fly general relativistic ray-tracing and subsequent convolutions of the relativistic effects with a complex underlying spectrum. This makes \texttt{reltrans} a good candidate for emulation, but also presents unique and constraining criteria for emulation: namely that it requires high precision (ideally less than $1\%$ error) and very fast computation (ideally less than 0.05 seconds per evaluation).

\subsection{reltrans}

\texttt{reltrans} assumes a lamppost point source corona emitting a Comptonization spectrum located above a black hole, which has a thermally emitting disk with a fixed aspect ratio of height over radius. The power law index of this spectrum is assumed to vary with time, motivated by empirical constraints from observations \citep{kotov2001x}. The spectral pivoting is implemented using a first order Taylor expansion to linearise the model. This emission from the corona travels not only to the observer, but also illuminates the disk. When emission from the corona hits the disk, it is reprocessed through scattering, absorption and emission by the material in the disk. This results in re-emission from the disk that is referred to as the reflection spectrum. The reflection spectrum has three key components that are apparent in the observed spectrum: the soft excess below 1 keV, the iron line complex around 6.4 keV and the Compton hump \citep{ross2005comprehensive,garcia2013x}.

This reflection spectrum is also affected by relativistic effects due to the corona and reflecting material being situated near the black hole \citep{fabian1989alignment}. These are modelled via general relativistic ray-tracing taking into consideration boosting from directed emission from the corona (rather than simply assuming isotropic emission), as well as blue- and red-shifting effects \citep[e.g.][]{2026MNRAS.545f1770B}. This results in relativistic smearing of the entire reflection spectrum, which is most apparent in the narrow line Fe K complex.

\begin{table}
    \centering
    \begin{tabular}{c|c}
        Parameter & Value \\
        \hline
        h & 6 $R_g$ \\
        a & 0.998 \\
        inc & $30.0^\circ$ \\
        $r_{inner}$ & 1 $R_{\mathrm{ISCO}}$ \\
        $r_{outer}$ & 2000 $R_g$ \\
        $z$ & 0.0 \\
        $\Gamma$ & 2.0 \\
        $\log\xi$ & 3.0 \\
        $A_{Fe}$ & 1.0 \\
        $\log N_e$ & 15.0 \\
        $kT_e$ & $60.0\mathrm{keV}$ \\
        $n_H$ & 0.0 \\
        boost & 1.0 \\
        Mass & $3\times10^6 M_\odot$
    \end{tabular}
    \caption{Default parameters used in Fig. \ref{fig:spectra_grid}.}
    \label{tab:spectra_grid}
\end{table}

We find it useful to visualise the behaviour of the model as a function of single parameters conditioned upon all other parameters to better understand the behaviour of the model. In Fig \ref{fig:spectra_grid}, we show how the spectrum evolves as a function of two different parameters of the model: the inner radius of the disk in the left panel and inclination in the right panel. All other parameters are fixed and shown in Table \ref{tab:spectra_grid}. The left panel shows that the spectrum becomes dimmer as the inner radius increases (as there is less disk to intercept radiation from the corona and re-emit it). We also observe that the lines present when the inner radius is large smear out and become considerably broader as the inner radius becomes smaller. This is because the light from the inner-most regions of the disk is both bolometrically brighter as well as most affected by relativistic distortion due to proximity to the black hole \citep[see e.g, the emissivity profiles in][]{2012MNRAS.424.1284W}. In the right hand panel, we instead show the spectral change with the inclination. The most obvious effect is that the blue wing of the iron line becomes bluer with higher inclination. This is because the maximum line of sight Doppler blueshift increases with increasing inclination. This maximum blueshift happens at about 10 $R_g$, since within that gravitational redshift starts to take over. There is a slight normalisation effect as the inclination approaches an edge-on system, where the disk emission is boosted towards the observer, both by gravitational lensing and the rotation of the disk. 

The non-linearity of these effects poses challenges for standard interpolation schemes in higher dimensions. In contrast, neural networks as universal function approximators are well suited to this type of problem. In BR25, we trained such an emulator for a particular flavour of \texttt{reltrans}, \texttt{RTDIST}. The emulator, \texttt{RTFAST}, was trained on the full original model, meaning that we generated pairs of input parameters and full output spectra from \texttt{RTDIST} and trained the neural network to learn the association between those. We also implemented Principal Component Analysis (PCA) to reduce the dimensionality of the output spectra and improve training performance. \changes{This makes \texttt{RTFAST} a special case of a deepONet (operator learning).} The resulting emulator is capable of reproducing the original model to a precision of $\lesssim 3\%$. In \texttt{reltrans}, the overall spectrum is made up of 2 components: the Comptonized spectrum continuum (\texttt{nthcomp}; \citealt{zdziarski1996broad}) and a relativistically convolved reflection spectrum (essentially equivalent to \texttt{relxill}; \citealt{dauser2013irradiation,dauser2016relativistic}). Much of the complexity of the reflection spectrum was swallowed up in the continuum and made emulation considerably easier overall. However, in actual observations, the reflection spectrum typically contributes $\approx10\%$ of the total spectrum. In constrast, \texttt{nthcomp} is trivial to compute (as it is only a relatively simple analytical function that approximates Comptonisation spectra) so a strategy of emulating \textit{only} the reflection spectrum to the same precision as the full spectrum would achieve a 10 times improvement in precision for "free" as well as reducing the amount of parameters (dimensions) to train on. Consequently, the aim of the current work is to train only on \texttt{reltrans}' reflection spectrum. Throughout the paper, we also discuss the limits of end-to-end emulation with similarly tight criteria as that imposed by our use-case.


\subsection{Training data generation}
\label{sec:trainingdata}

We generated 1.5 million pairs of parameter vectors, denoted as $\theta$, and corresponding model spectra $\hat{y}$ using the \texttt{reltransDCp} flavour of the \texttt{reltrans} v2.3.1 model family. These spectra were evaluated on a randomised logarithmically spaced energy grid, consisting of 1000 energies $E$ between 0.1 and 100 keV designed to cover the range of typical X-ray instruments used to study reflection in X-ray binaries. The energy bins are contiguous within this range and the resulting spectral values are divided by bin width such that the spectral values are in units of $\mathrm{photons}/\mathrm{s}/\mathrm{cm}^2/\mathrm{keV}$. \changes{As each model evaluation took approximately 0.3 seconds on a 12th Gen Intel(R) Core(TM) i7-12700K, the dataset took 125 CPU hours to compute and, after compression, a total of 17GB of storage.}

The model flux, parameters, and corresponding energy were then paired together, and the sets of ($\theta$, $E$, $y$)  randomised in training and validation. Parameter sets were generated using Latin hypercube sampling \citep{mckay2000comparison} with (flat) priors summarized in Table \ref{table:pars}\footnote{We additionally filter out parameter sets which fulfill the conditions: $\Gamma > 2.75$, $\log\xi > 4$ and $\log N_e > 17$. Under these conditions, the underlying \texttt{xillver} model no longer exhibits characteristic iron lines and acts as a purely Comptonized totally reflecting disk atmosphere.}. We fixed a number of parameters corresponding to effects that can be trivially included with the emulated model using analytical prescriptions, listed in table \ref{table:fixed}. All timing related parameters in the model were set to 0 as they are not relevant to time-averaged spectral modelling \footnote{This is also true for mass, which we set to $3\times10^6 M_\odot$.}. We set the density profile of the disk to be constant, and split the disk into 20 ionisation zones. We used a single angular zone for the disk (assuming that there is no asymmetric emission profile which is resultant from an isotropic corona). We split these sets of ($\theta$, $E$, $y$) \changes{-- based on parameter vectors $\theta$ --} 90:10 into training and test datasets. The training set was further divided 90:10 into a training and validation dataset. Once spectra were generated, we took the logarithm of the flux $y_i$ in each energy bin $E_i$ to obtain a new log-scaled quantity $\log(y_i)$. 

\texttt{reltrans} has slightly changed its definition of $kT_e$ in version v2.3 \citep[for version v2.2, see][]{ingram2022}, and is instead defined as the electron temperature in the source frame instead of the observer frame. This does not change the model physically, but brings \texttt{reltrans} in line with other models in the community and makes the inferred $kT_e$ more physically interpretable. There has also been a small change in how the relativistic kernel is convolved with the reflection spectrum in Fourier space to prevent numerical noise causing the relativistically convolved spectrum to be incorrect at high energies (>50keV) for large $\Gamma$ ($\geq 2$). This change is elaborated on in Appendix \ref{app:reltrans}.

\section{RTFAST2} \label{sec:sota}

Our goal is to emulate a computationally expensive, highly non-linear spectral model under two constraints that are unusually stringent for machine-learning surrogates: (i) the emulator must be fast enough to be used inside sampling-based inference, and (ii) it must be accurate at the percent level over a broad energy range. In this regime, architectural and loss-function choices are both equally important. They are often the difference between an emulator that is scientifically usable (and trustworthy) and one that yields confident but biased posteriors. 

\subsection{Techniques} \label{sec:emulator}

We utilise a variety of techniques that are still relatively uncommon in astronomical surrogate modelling: operator-learning, Fourier feature embeddings of the energy coordinate \citep{tancik2020fourierfeaturesletnetworks}, feature-wise linear modulation (FiLM) for conditioning on the physical parameters \citep{perez2018film}, trend heads to factor out smooth global behaviour, a loss aligned with percent-level relative error in linear flux, and a weak Sobolev-style regulariser to suppress non-physical oscillations \citep{czarnecki2017sobolevtrainingneuralnetworks}. In this subsection, we explain each of these choices and motivate why, taken together, they produce an emulator that is both fast and scientifically usable.

\subsubsection{Operator-learning}
The central design decision we adopt is to treat the spectrum as a \emph{function of energy conditioned on parameters}, that is as some $f(E;\theta)$, rather than as a fixed-length vector regression problem. This operator-learning formulation allows the emulator to learn interpolation across the energy axis. It furthermore becomes substantially easier to trade accuracy against evaluation speed in a controlled way by simply changing the resolution of the energy grid. Operator learning has recently begun to be adopted into neural network emulators in astronomy, for example in the chemistry of the interstellar medium \citep{branca2024}, the chemical evolution in cosmological simulations \citep{vanderbor2025}, disk-planetary interactions \citep{mao2023}, and to simulate radiative transfer in astrophysical environments \citep{rost2025}. 

A common baseline for spectral emulation is to train a network that maps parameters $\theta$ directly to a flux vector on a fixed grid,
\begin{equation}
\hat{\mathbf{y}}(\theta)\in\mathbb{R}^{M},
\end{equation}
where $M$ is the number of energy bins (here $M=1000$). While this is simple, it ties the emulator to a specific grid and encourages a ``memorise bin index'' failure mode when the spectrum contains narrow, shifting features.

Instead, we model the spectrum as a conditional function of a continuous coordinate $x$ (a transformed energy),
\begin{equation}
\hat{\ell}(x;\theta) \approx \ell(x;\theta),
\end{equation}
where $\ell \equiv \log_{10}(\hat{y})$\footnote{We use a $\:\hat{}\:$ to designate an emulator output.} and $x$ denotes the energy coordinate (we typically use $x=\log_{10}E$ and then normalise to a bounded interval for numerical stability). This formulation has two practical benefits. First, it naturally supports evaluation on arbitrary energy grids without re-training. Second, it enables a training strategy where the optimiser sees many different subsets of energy locations, which directly encourages interpolation along the energy axis rather than dependence on a fixed discretisation. In other words, this allows the emulator to be used with any X-ray instrument, rather than being constrained to a single instrument like that of BR25.

In our implementation this is realised by training either on full spectra (all $M$ points) or on randomly sampled subsets of points per spectrum. The latter reduces memory bandwidth and, importantly, makes ``learning the energy axis'' part of the task rather than an artefact of the training pipeline.

\subsubsection{Fourier embedding}

\begin{figure*}
    \centering
    \includegraphics[width=0.98\linewidth]{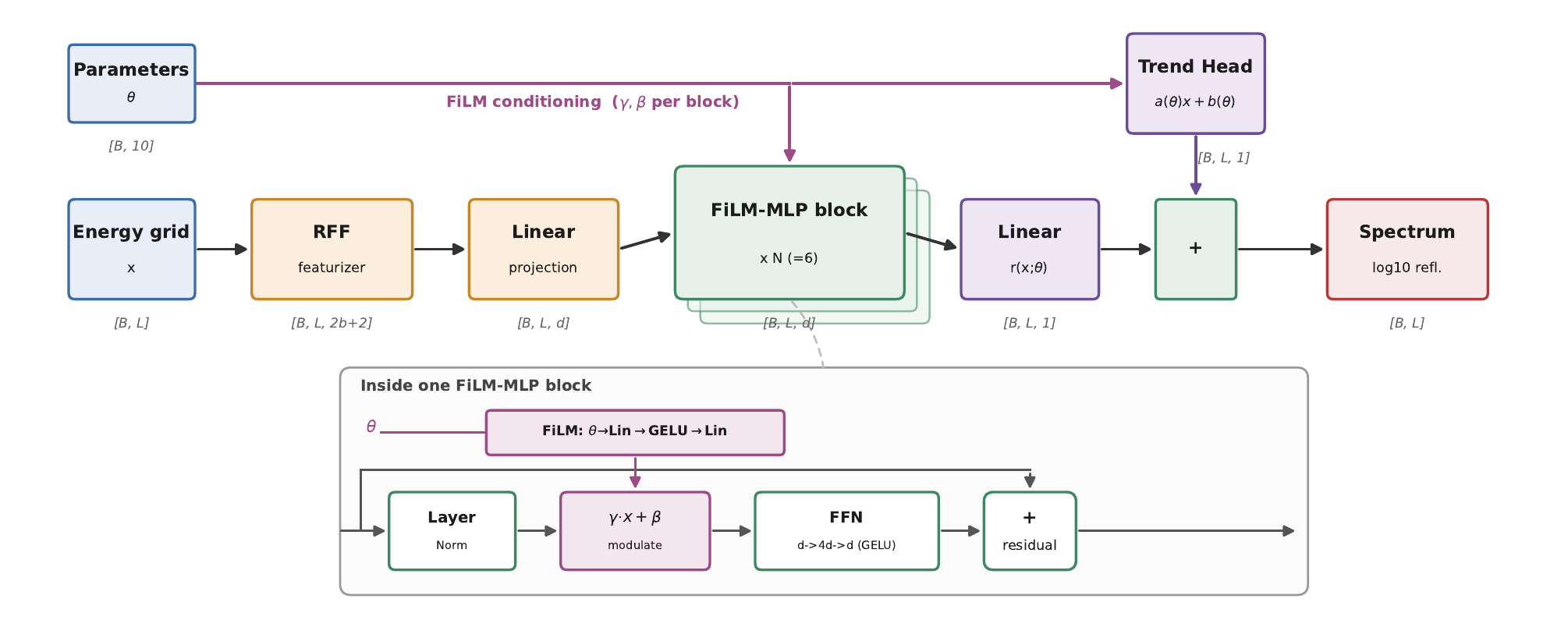}
    \caption{Visualisation of the emulator architecture. Input blocks are plotted in blue, featurizing blocks are plotted in orange, FiLM-MLP blocks are coloured in green, the head blocks (trend-heads and end of FiLM-MLP blocks) are plotted in purple, and the output spectrum block is plotted in red. The shape of the input shape of each block is shown below the block. B indicates the batch size, L indicates the length of the energy grid, N shows the number of FiLM-MLP blocks (6 in the case of the final network), d is the base width (376) of the feed-forward network (FFN), and b is the number of bands (256) in the random Fourier features (RFF). GELU is the Gaussian Error Linear Unit activation function \citep{hendrycks2016gaussian}.}
    \label{fig:architecture}
\end{figure*}

In BR25, we reduced the dimensionality of the output spectra by applying PCA to these spectra and transforming them into 200 PCA components, which were then used to train the neural network. However, this approach proved less fruitful here, largely due to the increased, non-linear complexity of the reflection spectrum component being emulating (for a discussion on PCA in this context, see Appendix \ref{sec:representation}). 

Reflection spectra contain both smooth components and sharp, localised structure (lines, edges, rapid curvature changes observable in Fig. \ref{fig:spectra_grid} and captured in Fig. \ref{fig:pca}), while standard multi-level-perceptrons (MLPs) often exhibit a ``spectral bias'' toward learning low-frequency structure first. This can lead to visually plausible but quantitatively poor reconstructions in precisely the regions that drive inference. Instead, to represent both smooth and sharply localised spectral structure, here we embed the energy coordinate using random Fourier features (RFF) before passing it to the network. Fourier features have been previously used across astronomy e.g.~ in TransformerPayne \citep{rozanski2025}, SpectraFM \citep{koblischke2024}, and FEDONet \citep{sojitra2026}. 

In its simplest form, the embedding maps a scalar $x$ to
\begin{equation}
\phi(x) = \big[ \cos(2\pi b_k x),\, \sin(2\pi b_k x) \big]_{k \in \{1, \dots, K\}},
\end{equation}
with frequencies $\{b_k\}$ spanning a range that captures the smallest feature scale we aim to reproduce at the target resolution. Intuitively, this makes oscillatory structure easy to represent: the network can form high-frequency behaviour via linear combinations of the embedded basis rather than requiring many layers to synthesise it from smooth activations.

A practical training detail is that very high-frequency embeddings can destabilise early optimisation. We therefore adopt a simple curriculum in which the effective Fourier frequencies are ramped up over the first part of training, allowing the model to first capture the global spectral shape and then progressively learn finer structure.

\subsubsection{Feature-wise linear Modulation} \label{sec:FiLM}

Within a residual neural network, blocks often begin to specialise in certain tasks. To guide this specialisation within the context of the physical model, we can condition the network on the physical parameters themselves. To do this, we use feature-wise linear modulation (FiLM). The mapping $\ell(x;\theta)$ is strongly non-linear in $\theta$ and highly sensitive: the effect of changing one physical parameter can depend on the values of several others. A naive conditioning approach is to concatenate $\theta$ to the embedded coordinate, i.e.\ learn $g([\phi(x),\theta])$. This works for some problems, but it tends to be inefficient when $\theta$ must modulate the behaviour of many intermediate features across the energy axis.

In a FiLM layer, intermediate activations $h$ (the normalized outputs of previous blocks and the projected RFF features) are transformed as
\begin{equation}
h' = \gamma(\theta)\odot h + \beta(\theta),
\end{equation}
where $\gamma(\theta)$ and $\beta(\theta)$ are learned functions of the physical parameters and $\odot$ denotes element-wise multiplication. This separates the roles of the two inputs: $x$ specifies \emph{where} in energy we evaluate the function, while $\theta$ specifies \emph{which} spectrum within the family we should produce. In practice, FiLM acts as a lightweight but expressive conditioning mechanism with minimal runtime overhead.

\subsubsection{Trend-head}

To factor out smooth global behaviour, we optionally include a small ``trend head'' that predicts coefficients $a(\theta)$ and $b(\theta)$ and adds a simple function of $x$,
\begin{equation} \label{eq:trend-head}
\hat{\ell}(x;\theta) = r(x;\theta) + a(\theta)\,x + b(\theta),
\end{equation}
where $r(x;\theta)$ is the main network output. Even after working in $\log_{10}$ flux, many spectra contain a smooth global trend (overall slope/curvature) plus local residual structure. This does not constrain the model to be linear in $x$; it merely makes it cheap to represent the dominant smooth component when present, leaving the residual network to focus on the hard, localised structure. Notably, as the emulator predicts the logarithm of the spectrum, this trend head is effectively a power law. We use this trend-head in the final emulator.

\subsubsection{Training loss}
Because the scientific requirement is naturally phrased in terms of percent-level errors, the training loss should reflect that requirement. Training purely with mean squared error in log space stabilises optimisation across large dynamic ranges, but it does not directly enforce small \emph{fractional} errors in linear flux.

We therefore train with a primary loss that is explicitly aligned with relative error in linear space. If the model predicts $\hat{\ell}$ and the target is $\ell$, then the corresponding relative error in linear flux is
\begin{equation} \label{eq:rel_loss}
\epsilon_{\rm rel}(x;\theta) = \left|\frac{\hat{y}-y}{y}\right|
= \left|10^{\hat{\ell}(x;\theta)-\ell(x;\theta)} - 1\right|.
\end{equation}
We minimise an average of $\epsilon_{\rm rel}$ (more precisely, a robust variant known as a Huber loss\footnote{Huber loss uses a squared term when the error falls below a certain $\delta$ (0.05 in our case) and a L1 term otherwise.} applied to $\epsilon_{\rm rel}$) over sampled energies. This objective has two desirable properties: it is naturally scale-free, and it treats over- and under-predictions symmetrically in linear space. In practice we sometimes include a small auxiliary stabilisation term in log space during early training to reduce the influence of occasional outliers while the model is still learning the coarse spectral shape.

For evaluation, we summarise performance in a way that matches the intended use in inference: for each test spectrum we compute the mean relative error across the $M$ energy bins, and report the fraction of test samples below fixed thresholds (e.g.\ $1\%$ and $10\%$). These threshold metrics are particularly informative because they reveal whether errors are broadly distributed or dominated by a minority of failure cases.

\subsubsection{Sobolev regulariser} \label{sec:sobolev}

To suppress small non-physical oscillations, we add a weak derivative-matching regulariser that penalises differences in $\partial \ell/\partial x$ between prediction and target. Such oscillations can contribute little to a standard pointwise loss yet still be undesirable in practice. A spectrum may therefore achieve a low average error because the oscillations are small in amplitude and can partially cancel across neighbouring bins, while still differing noticeably in local shape. In discrete form (using central differences), the term can be written as:

\begin{equation}
\mathcal{L} = \mathcal{L}_{\rm rel} + \lambda \left\|\frac{\Delta \hat{\ell}}{\Delta x} - \frac{\Delta \ell}{\Delta x}\right\|_2^2 .
\end{equation}

where $\mathcal{L}$ is the overall loss, $\mathcal{L}_{\rm rel}$ is the loss defined in Eq. \ref{eq:rel_loss}, and $\lambda$ is a weighting hyper-parameter. We apply this term in the transformed energy coordinate (typically $\log_{10}E$), where smoothness is more naturally defined over a broad energy range, and allows us to keep $\lambda$ small\footnote{on order of $10^{-3}$} so that it discourages artefacts rather than dominating the fit.

\subsection{Overview of emulator} \label{sec:emulator_sum}

Taken together, these choices constitute a ``state-of-the-art'' emulator for the present task: the model is formulated to interpolate along the energy axis, it is equipped with a representation that can express high-frequency structure, and it is trained with an objective that targets the percent-level requirements imposed by spectral inference. We visualise the architecture taken in our approach in Fig \ref{fig:architecture}. The figure is broken up into blocks of neural network components with the FiLM-MLP blocks repeating 6 times in our final neural network. 

The RFF is compromised of 256 bands producing a total of 512 fourier features, concatenated with the original x and a single vector equalling 1 (resulting in an even number shaped vector). These are then passed into the linear projection block which is a single linear layer that projects from the 514 input features to 376 features projection. These are then passed into the FiLM-MLP blocks which are residual blocks where the input of the previous block is added to the output of the block. The blocks themselves are compromised of a normalisation layer which are passed into an activation function with parameters defined by the FiLM layers as in Section \ref{sec:FiLM}. FiLM layers are made up of 2 linear layers of 256 hidden nodes with 752 outputted features (making up the 256 $\gamma$ and $\beta$ parameters for the activation function of each feature). The resulting output from the activation function is then passed into a feed forward neural network compromised of 2 layers: going from 376 features to 1504, passing through a GELU function, and then subsequently being reduced from 1504 features to 376 features again. Once passed through all 6 FiLM-MLP blocks, the 376 features are then compressed at the head of the networks into a single point prediction for $x$. This single point prediction is then added together with the predicted single value from the trend head (which is a 2 layer network that inputs the 10 parameters, expands to 128 hidden features and then predicts parameters $a$ and $b$ as in Eq. \ref{eq:trend-head}).

\changes{We report the evaluation cost of the final network. On the $4096$-bin \textit{NuSTAR} grid used in Section \ref{sec:ppc}, timed on a Macbook M1 Pro CPU over $128$ evaluations and taking the best of three repeats, \texttt{RTFAST2} takes $43$ ms per evaluation called one spectrum at a time, $28.6$ ms per evaluation when the whole ensemble is passed through in a single batched forward pass on 8 threads, and $14.5$ ms per evaluation across a 12 process pool. Evaluated in the same configuration, \texttt{reltransDCp} takes $132$ ms serially and $26.5$ ms across the same 12 process pool. The emulator meets the $\leq0.05$ s per evaluation criterion set out in Section \ref{sec:reltrans} without parallelisation, and is a factor of three faster than the original model called the same way. \texttt{reltransDCp} parallelises better, as its natural mode is one call per parameter vector whereas a pool forces \texttt{RTFAST2} to give up batching, so at 12 processes the gap narrows to a factor of two. Folding the result through the response adds $0.7$ ms per spectrum when done as a single batched matrix product. As discussed in Section \ref{sec:ppc}, the larger gain in practice is in the number of steps required to converge rather than in the cost of a step. Notably, due to the operator learning approach taken in \texttt{RTFAST2}, computation time scales directly with grid size so the emulator is a greater improvement for coarser grids than the above comparison.}

\changes{The trained network is $36$ MB of weights, which is a significant reduction in size to the approximately $7$ GB of tabulated \texttt{xillver} grids that \texttt{reltransDCp} interpolates to compute the rest-frame reflection spectrum. The emulator can be distributed with the model and held in memory.}

\subsection{Ablation studies}

\begin{table*}
\centering
\renewcommand{\arraystretch}{1.2} 
\begin{tabular}{@{} l r r c r r @{}} 
\toprule
& \multicolumn{2}{c}{Strict Threshold ($\epsilon_{\rm rel} \le 0.01$)} & \phantom{a} & \multicolumn{2}{c}{Loose Threshold ($\epsilon_{\rm rel} \le 0.10$)} \\
\cmidrule{2-3} \cmidrule{5-6}
Architecture Configuration & Yield (\%) & Drop ($\Delta$) && Yield (\%) & Drop ($\Delta$) \\
\midrule
Base Model                 & 85.24 & ---   && 99.99 & ---  \\
\quad w/o Sobolev Regularization & 84.61 & 0.74  && 99.99 & 0.00 \\
\quad w/o Trend Head       & 85.01 & 0.27  && 99.99 & 0.00 \\
\quad w/o FiLM Conditioning& 76.94 & 9.74  && 99.99 & 0.00 \\
\quad w/o Fourier Features & 80.29 & 5.81  && 99.99 & 0.00 \\
\bottomrule
\end{tabular}
\caption{Ablation study of model components. Performance is measured by the percentage of test samples falling within a specified relative error threshold ($\epsilon_{\rm rel}$). The drop ($\Delta$) indicates the absolute percentage point decrease in performance compared to the Base model.}
\label{tab:ablation_results_tight}
\end{table*}

Because the final accuracy depends on several interacting choices, we performed targeted ablation studies in which we removed one component at a time from the base configuration, allowing us to understand the magnitude of each technique's effect on the final result. The results (Table~\ref{tab:ablation_results_tight}) show that conditioning and coordinate embeddings matter most: removing FiLM conditioning or Fourier features produces the largest degradation in strict ($1\%$) accuracy, whereas removing trend decomposition or Sobolev regularisation produces smaller changes. These outcomes support the interpretation that the hardest part of this problem is representing strongly parameter-dependent, localised structure in the spectrum, rather than capturing global trends. This isn't out-of-line with expectations: the global trend in a reflection spectrum is simply a power-law, while the non-linear behaviour is a result of relativistic effects dependent on the black hole properties and the location of the observer with respect to the system.

\subsection{Performance Scaling}

\begin{table}
\centering

\renewcommand{\arraystretch}{1.2}
\begin{tabular}{@{} l r r @{}}
\toprule
Model Size & \multicolumn{2}{c}{Yield (\%)} \\
\cmidrule{2-3}
(Parameters) & $\epsilon_{\rm rel} \le 0.01$ & $\epsilon_{\rm rel} \le 0.10$ \\
\midrule
16K  & 49.77 & 93.49 \\
512K & 75.47 & 99.90 \\
8M   & 85.24 & 99.90 \\
32M  & 85.02 & 99.90 \\
\bottomrule
\end{tabular}
\caption{Model scaling results. Performance is measured by the percentage of test samples falling within strict ($1\%$) and loose ($10\%$) relative error thresholds ($\epsilon_{\rm rel}$).}
\label{tab:model_scaling_raw}
\end{table}

Beyond architecture, an equally practical question is how emulator quality scales with the amount of training data and with model capacity. In this setting, generating additional training spectra is itself computationally expensive, so the relevant quantity is not only the lowest attainable error, but also the rate at which performance improves as more data are added. A useful way to present this is to fix the optimisation and hyper-parameter setup, vary the size of the training set, and plot the resulting training loss alongside a test-set metric defined in linear space, such as the mean relative error or the fraction of spectra below fixed error thresholds. Such a comparison shows whether the emulator continues to benefit from additional data or whether improvements begin to saturate, indicating that further pre-generation of spectra may no longer be worthwhile.

The same logic applies to model capacity. By varying network width or depth at fixed data volume, we can compare two different uses of computational budget: increasing the size of the emulator or generating more training spectra. The better choice need not be the same in every regime. In the low-data regime, a modest increase in capacity may extract more information from the available samples, whereas once the model is sufficiently expressive, further gains may depend more strongly on the breadth of the training set. The purpose of this subsection is therefore to identify where data scaling remains worthwhile, where model scaling is more effective, and where both begin to exhibit diminishing returns. These trade-offs are especially important for surrogate modelling, where the cost of producing new training data can be comparable to, or larger than, the cost of training the emulator itself.

In Table \ref{tab:model_scaling_raw}, we see that an increase in network parameters (by either increasing the number of layers, the number of nodes in each layer, or both) yields significant gains in performance in particular for reducing outlying poor performance but eventually saturating such that there is no longer a need to increase model size past 8M parameters\footnote{This is the final network size as described in Section \ref{sec:emulator_sum}.}. We found in our experiments that test loss improved logarithmically with respect to data size, with test loss approximately halving with every 10 times increase in training data. While the general principle of generalisation improving logarithmically with data size tends to hold across problems, the exact slope of improvement tends to be problem dependent and is thus always worth assessing in terms of trade-off for each individual problem. 

\section{Performance in Astrophysical Modelling} \label{sec:ppc}

\begin{figure}
    \centering
    \includegraphics[width=\linewidth]{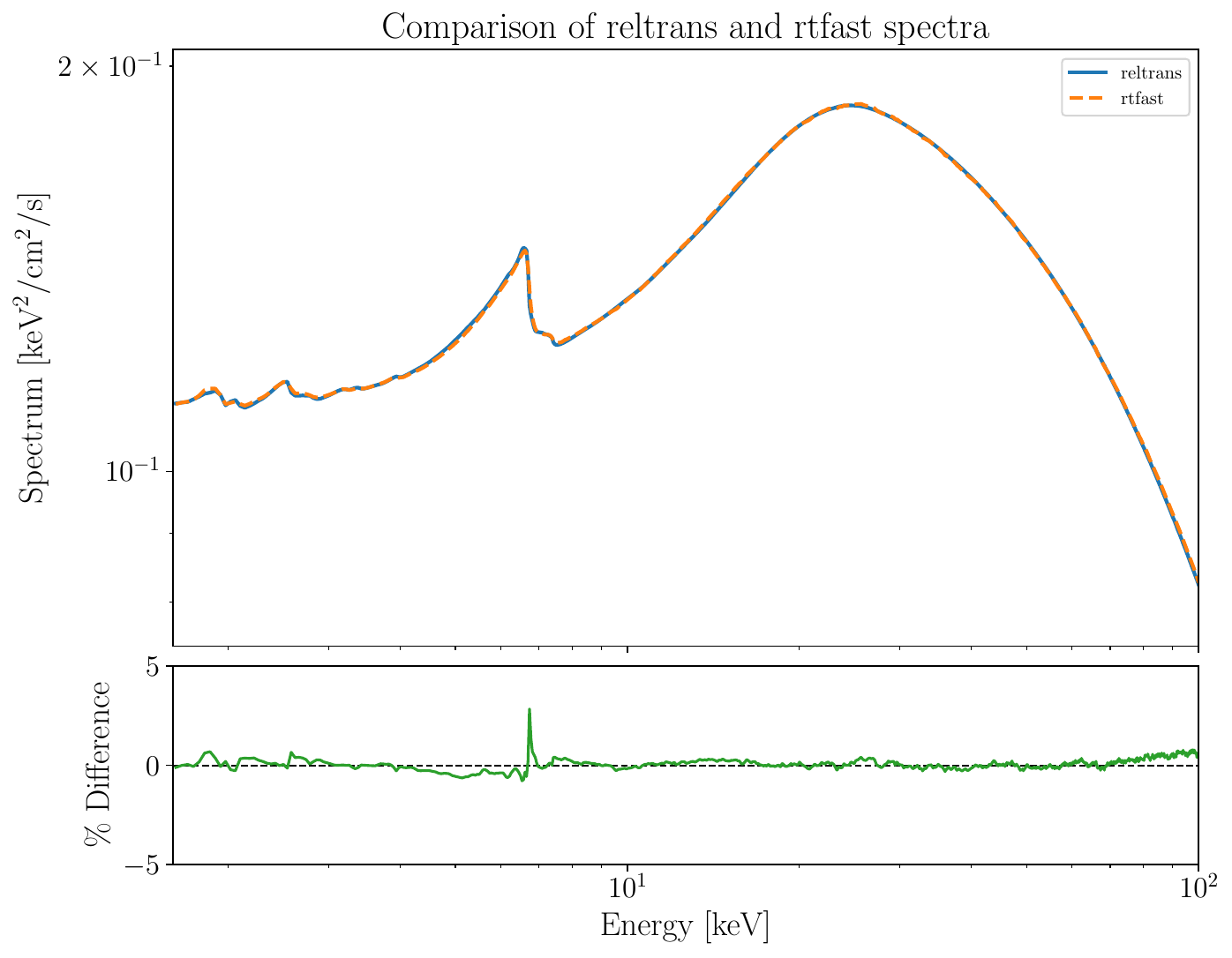}
    \caption{Top panel: the \texttt{reltrans} (blue solid) and \texttt{RTFAST2} (orange dashed) spectra  in $keV/cm^2/s$ units. Bottom panel: The fractional difference between \texttt{RTFAST2} and \texttt{reltrans}. A value of 1 corresponds to a 1\% offset of the emulated spectrum compared to the original \texttt{reltrans} value.}
    \label{fig:spectra_compare}
\end{figure}

This section will now measure the emulator's performance in the practical task it was designed to do: model astrophysical spectra of accreting black holes. While error values and statistics are useful for understanding the performance of the emulator in a more machine-learning context, in practice we need to ensure the emulator is capable of inferring parameters that are biased only within tolerance (typically, of the order of magnitude of the statistical and systematic errors present in the data). 

We first compare \texttt{reltrans} to \texttt{RTFAST2} directly. In Fig \ref{fig:spectra_compare}, we plot the \texttt{reltrans} spectrum and the \texttt{RTFAST2} spectrum for the parameters shown in Table \ref{tab:simulated_pars} as a direct comparison. One can see that the percentage difference is well below $10\%$. There is a small spike in error on order of a few $\%$ near the iron line but there are otherwise no systematic errors.

\begin{figure}
    \centering
    \includegraphics[width=\linewidth]{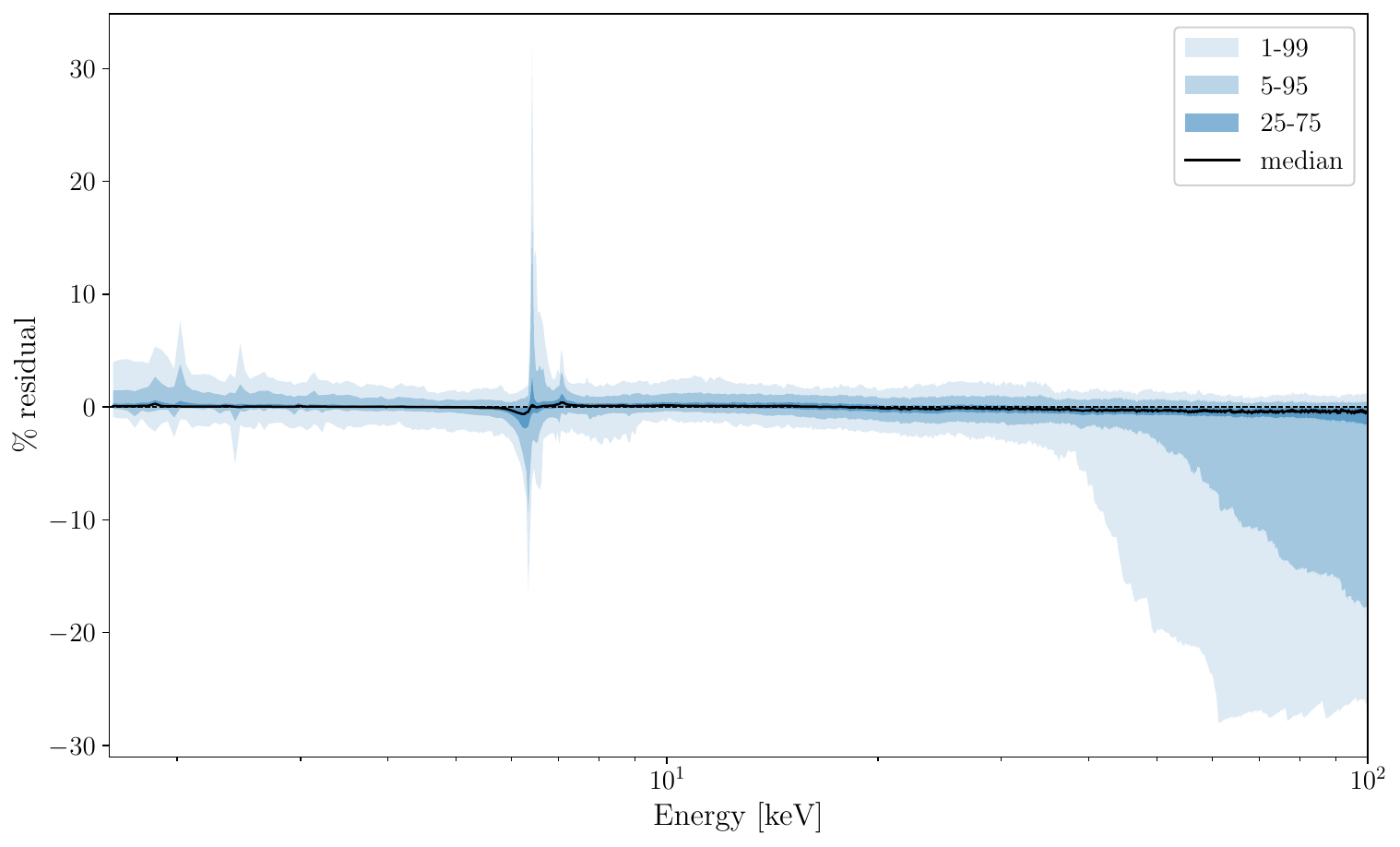}
    \caption{Distribution of the percentage residual difference between \texttt{RTFAST2} and \texttt{reltrans} as a function of energy. The shade of region indicates the $25-75\%$, $5-95\%$, and $1-99\%$ region in increasing lightness respectively.}
    \label{fig:error_distribution}
\end{figure}

While a single spectrum comparison is useful, we also need to understand the overall performance as a function of energy across the spectrum. If there is a bias in the emulator with respect to energy (as was somewhat the case in BR25), this points to potential sources of bias in using the emulator to infer astrophyiscal parameters. It also provides a useful diagnostic in understanding how model architecture choices affect bias in our emulation. We plot the distribution of percentage residuals between \texttt{reltrans} and \texttt{RTFAST2} in Fig \ref{fig:error_distribution} (essentially a summary of the bottom panel of Fig \ref{fig:spectra_compare} for all spectra in the test data set). When comparing to the analogous Fig 11 in BR25, it is obvious that the distribution for \texttt{RTFAST2} is much tighter, showing a clear improvement. There is however a very narrow spike at just above and just below 6.4keV : this is due to the fast changes in the reflection spectrum at these energies. The weak derivative-matching regulariser described in Section \ref{sec:sota} is able to reduce this inaccuracy somewhat during training but not completely. We also see a larger spread in errors at high energies ($\geq60$ keV). This is largely not a practical concern: there is very little emission at high energies and is not constraining for parameters of interest such as spin and inner radius of the disk.

Finally, we test the emulator's performance in an astrophysical setting by embedding it in a Bayesian inference procedure. In this case, we want to see how the emulator expects to perform in a "real world" setting. We do however want some control over the scenario, so we simulate an observation using \texttt{reltrans}. We then sample the posterior probability distribution of the parameters using Markov Chain Monte Carlo as implemented in \texttt{emcee }\citep{foreman2013emcee}\changes{, an implementation of the affine-invariant ensemble sampler of \citet{goodman2010}}, using the trained emulator as drop-in replacement for \texttt{reltrans} in the modelling. Ideally, the inferred posterior density should contain the true set of parameters; this implies that any bias in the emulator is minimal.

\begin{figure*}
    \centering
    \includegraphics[width=\linewidth]{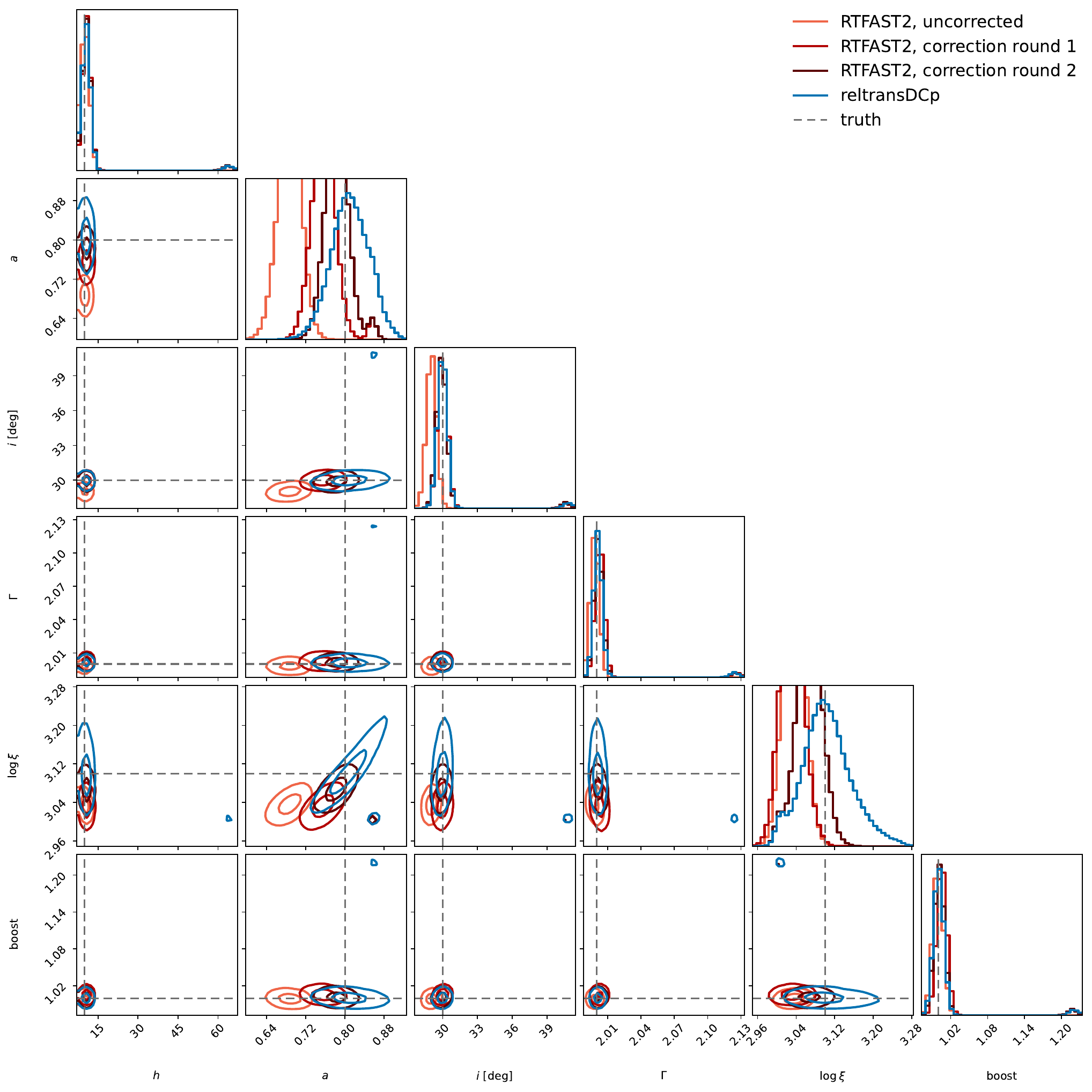}
    \caption{\changes{Posteriors on simulated \texttt{reltrans} data. We show \texttt{RTFAST2}
    uncorrected, after each of the two rounds of the multiplicative correction of
    Section \ref{sec:correction}, and a reference inference run with \texttt{reltransDCp} itself on
    the same simulated data. The true values used to generate the simulated data are marked by dashed grey lines.}}
    \label{fig:posteriors}
\end{figure*}

\begin{figure*}
    \centering
    \includegraphics[width=\linewidth]{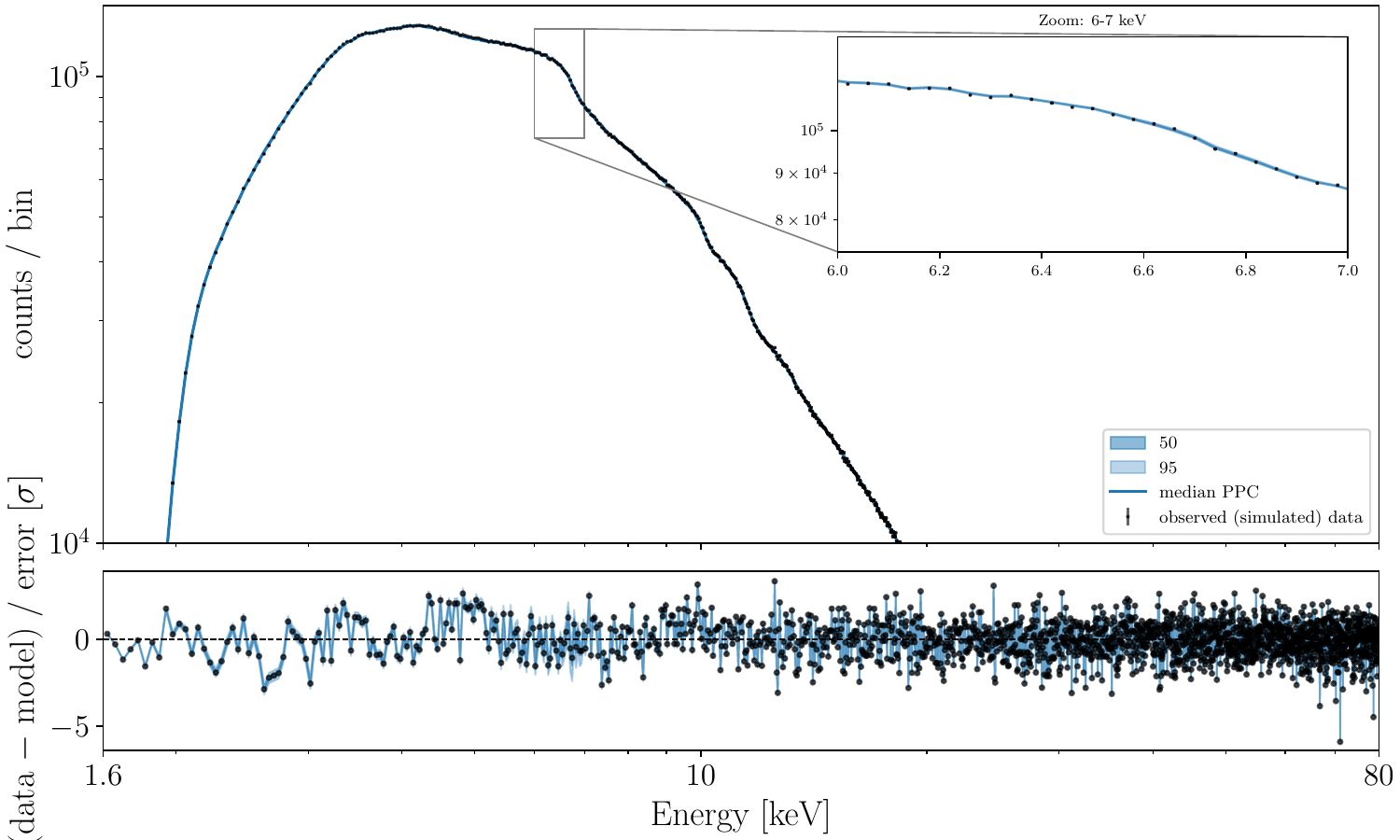}
    \caption{Top panel: Plot of the distribution of 200 model evaluations drawn from the posterior, compared to the simulated data. A zoomed in-set panel shows the posteriors draw distribution about the relativistic Fe-K line. While we plot the $25-75\%$ (dark blue shaded) and $2.5-97.5\%$ (light blue shaded) bands as filled in bands, they are not visible to the naked eye in the full plot and are tightly distributed in the zoomed plot, showing the tight constraints. The median posterior predictive draw is plotted as a solid blue line that passes close to through the data. Bottom panel: We plot the distribution of residuals (coloured same as the top panel) as a function of energy. The median residuals are plotted as black dots.}
    \label{fig:ppc}
\end{figure*}

\begin{table}
    \centering
    \begin{tabular}{c|c|c}
        Parameter & Value & Fixed \\
        \hline
        h & 10.0 & N \\
        a & 0.8 & N \\
        inc & 30 & N \\
        $R_{\text{inner}}$ & -1 & Y \\
        $R_{\text{inner}}$ & 400 & Y \\
        $z$ & 0.0 & Y \\
        $\Gamma$ & 2.0 & N \\
        $\log \xi$ & 3.1 & N \\
        $A_{Fe}$ & 1.0 & Y \\
        $\log N_e$ & 15.0 & Y \\
        kTe & 100.0 & Y \\
        nH & 0.1 & Y \\
        boost & 1.0 & N \\
        norm & 0.002 & Y \\
    \end{tabular}
    \caption{List of \texttt{reltrans} parameters used in generating a simulated spectrum. We only list parameters relevant to the time-averaged spectrum. We indicate the parameters kept fixed in the right hand column. The explanation for each parameter is listed in Table \ref{table:pars} and \ref{table:fixed}.}
    \label{tab:simulated_pars}
\end{table}

We choose a set of \texttt{reltrans} parameters that represent a realistic example of the type of scenario that this model would be used (in this case an AGN; parameters listed in Table  \ref{tab:simulated_pars}). We simulate a spectrum by convolving the \texttt{reltrans} model output with the FPMA NuSTAR response with an effective exposure of 100 ks. Then, we take a Poisson realisation of the convolved model output to simulate observation of real data.

We then perform inference by embedding \texttt{RTFAST2} into a Bayesian model and sample the posterior via MCMC \citep{metropolis1953equation,10.1093/biomet/57.1.97}. We utilize a Gaussian log-likelihood with additional systematic error added in quadrature to model the inaccuracies within the model\footnote{We model the inaccuracies within the model as a flat $0.5\%$ systematic errors across all energies. \changes{This value is motivated by the median emulator error: the scan described in Appendix \ref{app:failures} gives a median mean relative error of $0.43\%$ per spectrum, or $0.46\%$ when weighted by flux. The error is not uniform, either across parameter space (Appendix \ref{app:failures}) or across energy (Fig. \ref{fig:error_distribution}), so a single number is an approximation. We address the energy dependence directly in Section \ref{sec:correction}. This size of error is also consistent with the systematic error often adopted in NuSTAR analyses, thus reproducing posteriors more similar to what we see in actual analysis.}}. We model the prior distribution of each parameter as truncated Gaussian distributions centered on the truth with a full width half maximum of $20\%$ of the parameter's true value. We utilized 32 walkers and ran the MCMC for a maximum of 50,000 steps per walker\footnote{Each MCMC iteration (32 walker likelihood evaluations) took approximately 1 second on a Macbook M1 pro CPU. This will be faster on a GPU and scale better than $O(n)$ with more walkers due to vectorization gains.}, checking for convergence (number of steps exceed 50 times the estimated auto-correlation length in line with recommendations of \citealt{foreman2013emcee}) every 100 steps. In reality, the chain converged after 5600 steps. This is considerably faster than \texttt{reltransDCp} which usually takes on order of $10^6-10^8$ steps to converge on actual data and approximately $2\times10^4$ steps on this simulation\footnote{This is also true even when no systematic errors are added: MCMC converged in approximately 16000 steps without systematics accounting for emulator error} (further discussion in Appendix \ref{sec:chi2}). This is a somewhat unexpected (but welcome) gain over the original model and represents a larger speed-up \changes{in inference} that cannot be compared by the raw \changes{computational} speed of \texttt{RTFAST2} and \texttt{reltrans} alone.

The resulting posteriors are plotted in orange in Fig.\ref{fig:posteriors}. The trace plots can be found in Appendix \ref{sec:MCMC}. Unlike in BR25, we successfully recover the truth, with most parameters falling within 1 or 2 sigma posteriors in the 1-D posteriors. A notable exception is spin. This however is somewhat unsurprising: spin is notoriously difficult to measure in these systems and the results are affected by multiple parameters simultaneously. In addition, \texttt{RTFAST2}'s tendency to have a slightly more blue-shifted iron line (see Fig. \ref{fig:error_distribution}) than the ground truth leads to a under-prediction of the inclination (which is correlated with spin). \changes{Quantitatively, spin, inclination and $\log\xi$ are biased by $-5.1\sigma$, $-4.0\sigma$ and $-3.2\sigma$ respectively, while $h$, $\Gamma$ and the reflection normalisation all lie within $1.5\sigma$ (Table \ref{tab:correction}). In Section \ref{sec:correction}, we show that this residual bias can be removed by measuring the emulator error directly and applying it as a multiplicative correction.} 

As part of our analysis, we also plot posterior predictions in Fig. \ref{fig:ppc}. We randomly sample the posterior 200 times and then plot the distributions of the resulting model evaluations against the simulated data. We see that the distribution tightly fits the simulated data in the top panel as expected. In a zoomed-in panel showing the relativistically smeared Fe-K line, we see that the posterior draws are tightly constrained about the simulated data but are slightly broader than the rest of the spectrum, reflecting the wide posteriors in spin and $\log\xi$. The bottom panel shows the distribution of residuals for the same model evaluations. We see a somewhat broad distribution of possible model evaluations at lower energies (somewhat reflecting the larger systematic at below 5keV) with the median posterior draw residuals scattering below and above the zero line and consistently near zero above 15keV. 

The recovery of the truth in Fig. \ref{fig:posteriors} and the shared minimum chi-squared between the emulator and original model in Appendix \ref{sec:chi2} better reflects the capability of the emulator in this case.

\changes{\subsection{Correcting residual emulator bias} \label{sec:correction}}

\changes{The residual bias discussed above arises from a systematic, energy-dependent discrepancy between \texttt{RTFAST2} and \texttt{reltransDCp}. This is an unusually favourable case of model misspecification: because both models can be evaluated at the same parameter vector, the discrepancy can be measured directly rather than having to be absorbed by a nuisance component in the likelihood. We therefore measure the error itself and apply it as a multiplicative correction to the emulator output.}

\changes{For a parameter vector $\theta$, we define the log-ratio residual on the model energy grid, before the instrument response is applied,}
\begin{equation}
    d(E;\theta) = \ln S_{\texttt{reltransDCp}}(E;\theta) - \ln S_{\texttt{RTFAST2}}(E;\theta),
    \label{eq:logratio}
\end{equation}
\changes{where $S$ is the reflection spectrum returned by each model. We work with the log-ratio because the emulator error is fractional rather than additive, in the same sense as the training loss of Eq. \ref{eq:rel_loss}, and because it is scale-free: the overall model normalisation cancels, so the correction cannot absorb it. Measuring $d$ before the convolution with the response also keeps the correction instrument-independent, in keeping with the resolution-independent design of the emulator, so that the same correction can be applied with any response matrix.}

\begin{figure*}
    \centering
    \includegraphics[width=0.8\linewidth]{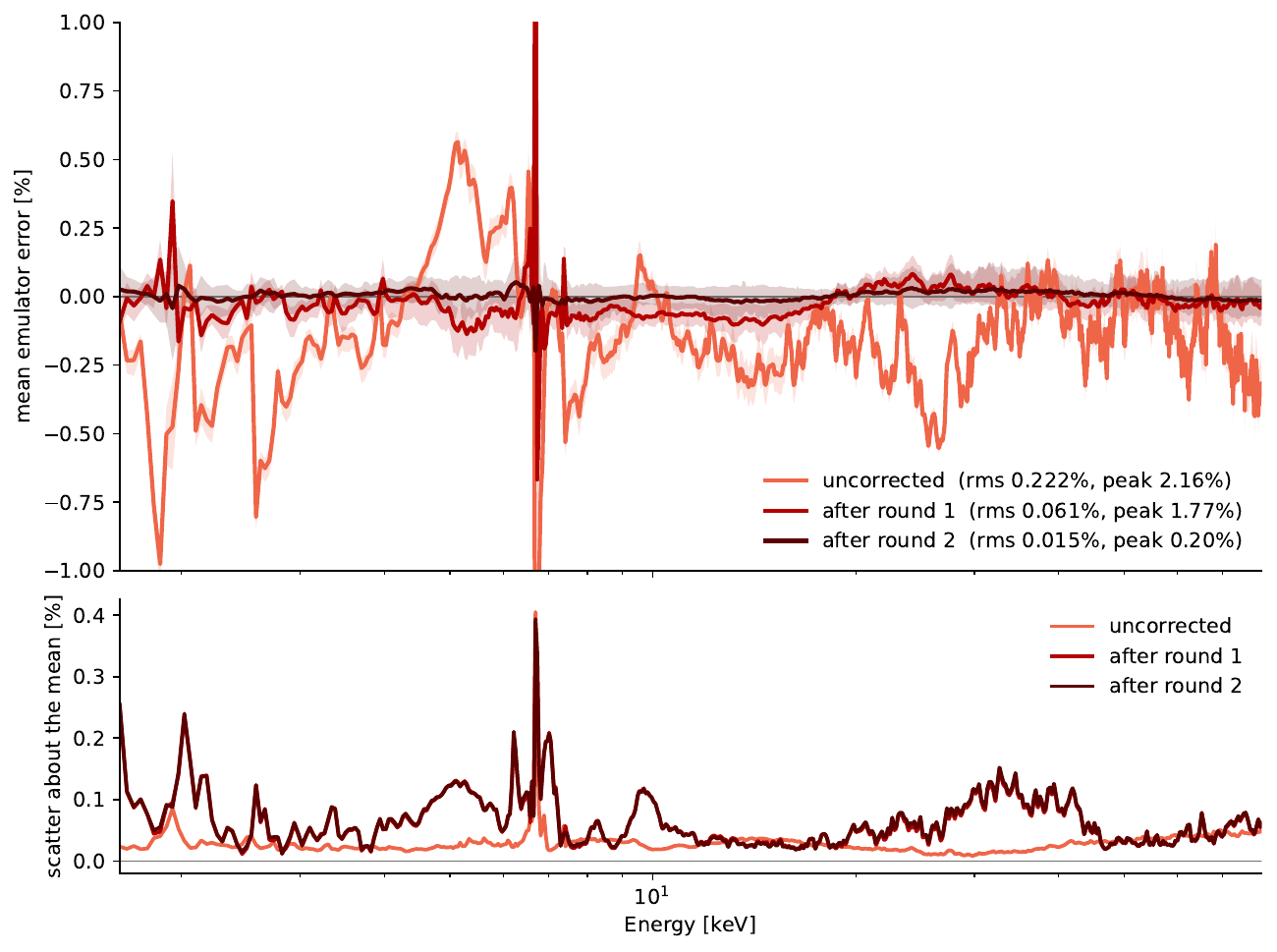}
        \caption{\changes{Emulator error before the correction and after each round. Top: the mean error, with the shaded band giving the $95\%$ range across draws. The broad continuum structure is removed after the first round, leaving a narrow residual at the Fe K edge. Bottom: the parameter-dependent scatter about the mean. Because each round samples a different region of parameter space, the scatter is not directly comparable between curves and its apparent increase does not indicate worse emulation.}}
    \label{fig:emuerr_decomp}
\end{figure*}

\begin{figure}
    \centering
    \includegraphics[width=\linewidth]{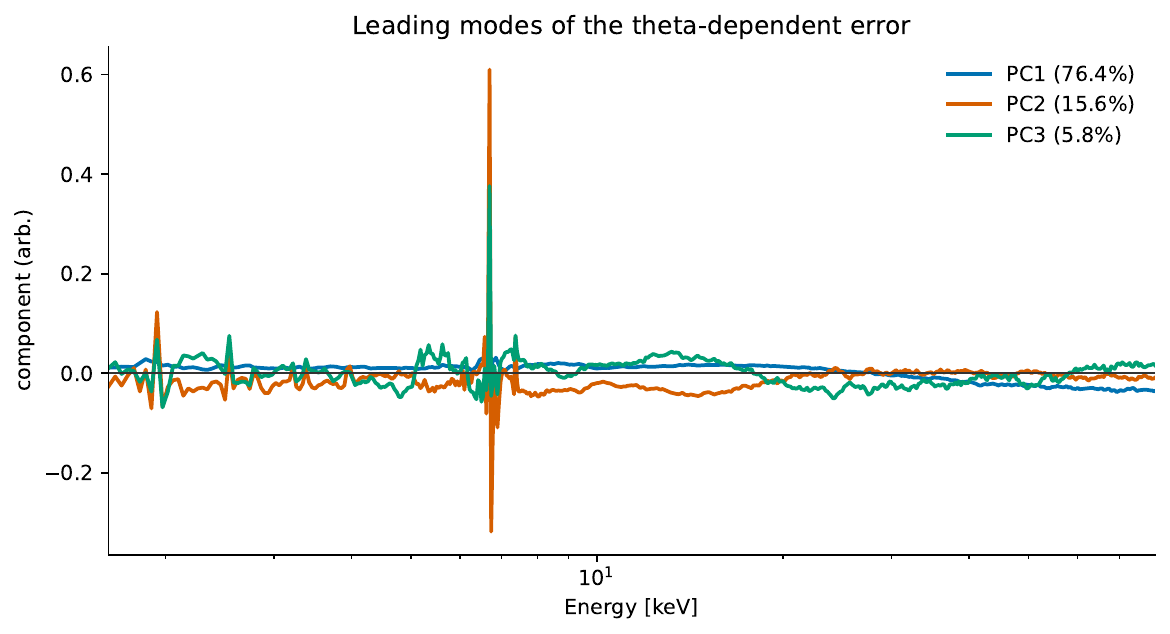}
    \caption{\changes{Leading principal components of the parameter-dependent error $\delta$. The second and third components are sharply localised at the Fe K edge.}}
    \label{fig:emuerr_pca}
\end{figure}

\changes{We drew 500 parameter vectors from the posterior of the inference described above, evaluated both models at each, and then split $d$ into a fixed component and a parameter-dependent remainder,}
\begin{equation}
    d(E;\theta) = \hat{d}(E) + \delta(E;\theta), \qquad
    \hat{d}(E) = \left\langle d(E;\theta) \right\rangle_{\theta},
    \label{eq:decomp}
\end{equation}
\changes{where $\left\langle \cdot \right\rangle_{\theta}$ denotes the average over the 500 draws. We exclude channels above 80 keV, where the simulated data contain no counts. We find that the fixed component dominates: $\hat{d}$ carries $97.7\%$ of the residual variance, with an rms of $0.222\%$ and a peak of $2.16\%$ at the Fe K edge, while $\delta$ has an rms of only $0.034\%$ (Fig. \ref{fig:emuerr_decomp}). A PCA decomposition of $\delta$ places $76.4\%$, $15.6\%$ and $5.8\%$ of the remaining variance in the first three components, with the second and third sharply localised at the Fe K edge (Fig. \ref{fig:emuerr_pca}). The parameter dependence that remains is therefore driven by the position and width of the relativistic iron line, and hence by spin and inclination.}

\begin{table*}
\centering
\caption{\changes{Posterior recovery before and after the multiplicative emulator-error correction. Each cell gives the posterior median, with the bias $(\mathrm{median}-\mathrm{truth})/\sigma$ in parentheses, where $\sigma$ is half the 16th--84th percentile range of the marginal posterior. The \texttt{reltransDCp} column is the reference inference run with the original model shown in Fig. \ref{fig:posteriors}.}}
\label{tab:correction}
\begin{tabular}{lrrrrr}
\hline
 & & \multicolumn{3}{c}{\texttt{RTFAST2}} & \\
\cline{3-5}
Parameter & Truth & Uncorrected & Correction round 1 & Correction round 2 & \texttt{reltransDCp} \\
\hline
$h$ [$R_{g}$] & 10.000 & 10.0119 ($+0.1\sigma$) & 10.1866 ($+1.9\sigma$) & 10.1441 ($+1.4\sigma$) & 10.0866 ($+0.9\sigma$) \\
$a$           &  0.800 &  0.6865 ($-5.1\sigma$) &  0.7566 ($-2.0\sigma$) &  0.7832 ($-0.8\sigma$) &  0.8102 ($+0.3\sigma$) \\
$i$ [deg]     & 30.000 & 29.0288 ($-4.0\sigma$) & 29.9984 ($-0.0\sigma$) & 29.9062 ($-0.4\sigma$) & 29.9865 ($-0.1\sigma$) \\
$\Gamma$      &  2.000 &  1.9983 ($-1.5\sigma$) &  2.0027 ($+2.1\sigma$) &  2.0017 ($+1.4\sigma$) &  2.0011 ($+1.0\sigma$) \\
$\log\xi$     &  3.100 &  3.0357 ($-3.2\sigma$) &  3.0311 ($-3.0\sigma$) &  3.0665 ($-1.4\sigma$) &  3.1008 ($+0.0\sigma$) \\
boost         &  1.000 &  0.9993 ($-0.1\sigma$) &  1.0059 ($+1.2\sigma$) &  1.0020 ($+0.4\sigma$) &  1.0006 ($+0.1\sigma$) \\
\hline
\end{tabular}
\end{table*}

\changes{Applying the fixed component alone, again before the convolution with the response, gives the corrected model}
\begin{equation}
    S_{\mathrm{corr}}(E;\theta) = S_{\texttt{RTFAST2}}(E;\theta)\,\exp\!\left[\hat{d}(E)\right].
    \label{eq:correction}
\end{equation}
\changes{As $\hat{d}(E)$ is a fixed vector on the model energy grid, the correction in Eq. \ref{eq:correction} introduces no free parameters, requires no re-training of the network and adds negligible cost to a likelihood evaluation. This is a cheap correction as it only requires the 500 \texttt{reltransDCp} evaluations used to calibrate it, a small fraction of the model evaluations required to sample the posterior.}

\changes{Because $\hat{d}$ is measured over the region of parameter space sampled by a run that is itself biased, we iterate. We re-run the inference with the first correction applied, draw from the resulting posterior, and measure the log-ratio of \texttt{reltransDCp} against the corrected emulator; the total correction is then the product of the two, $\exp[\hat{d}_1 + \hat{d}_2]$, which is a single vector applied once at run time. Since $\hat{d}_1$ is a constant vector it can be taken out of the average, so this total is identical to the absolute log-ratio measured over the new region. We work with the increments because $\hat{d}_2$ then measures the error left after the first round, which is what indicates whether a further round is needed. In our case, two rounds were sufficient: the second increment had an rms of $0.061\%$, $27\%$ of the first, and reduced the per-draw $\chi^2$ of the discrepancy against the error bars of the data from $6.92$ to $0.86$.}

\changes{The resulting parameter recovery is listed in Table \ref{tab:correction} and the posteriors are plotted in Fig. \ref{fig:posteriors}. The bias in spin falls from $-5.1\sigma$ to $-0.8\sigma$, inclination from $-4.0\sigma$ to $-0.4\sigma$ and $\log \xi$ from $-3.2\sigma$ to $-1.4\sigma$; in absolute terms, the recovered spin moves from $a=0.687$ to $a=0.783$ against a true value of $a=0.8$. It is worth noting that the appropriate benchmark here is agreement with \texttt{reltransDCp} rather than with the data truth, as the reference run carries its own offset from the truth for this particular noise realisation. On that basis, all six free parameters agree after correction.}

\changes{It should be acknowledged that this is a mean correction: the parameter-dependent component $\delta$ is not removed. Capturing it would require a parameter-dependent error model, for example a principal component decomposition with the coefficients regressed on $\theta$ together with a covariance term in the likelihood. Since the residual $\delta$ is $0.034\%$ rms, more than an order of magnitude below the $0.5\%$ systematic already adopted, we do not find this necessary here. The correction also does not improve the emulator itself; it removes the systematic component of its error at the point of use, and the residual spike at the Fe K edge remains the clearest target for future training improvements.}
\vspace{0.25cm}

\section{Discussion}

We have designed and implemented \texttt{RTFAST2}, utilizing a series of state-of-the-art machine learning methods to achieve a high performance, resolution-independent emulator for the relativistically convolved reflection spectrum of \texttt{reltransDCp} flavour of the \texttt{reltrans} model family. The emulator is integrated into the original physical model, takes a modular approach and shows only a small bias when used to model simulations generated using \texttt{reltrans} itself. We achieve low overall errors over the parameter space, \changes{including a relative error of less than $1\%$ and $10\%$ on $85\%$ and $99.9\%$ of energy bins respectively (we stress that these fractions are of the energy space, not of test spectra; the corresponding per-spectrum figures are given in Appendix \ref{app:failures}).} This high performance on a highly non-linear model with a large and sparse parameter space shows the strength of emulation for complex models.

We show that these improvements in performance over BR25 enable a significant reduction in inference bias and enable \texttt{RTFAST2} to be a trust-worthy part of the inference pipeline of measuring black hole properties in the future. The approach taken in this paper also makes the emulator far more flexible than its previous iteration, enabling the use of the model with data from any X-ray instrument within the energy range the model was trained in, instead of being limited to a single instrument.

The rest of this section is dedicated to discussing our results further and how we believe that it should inform future work on emulators, both within \texttt{reltrans} and more broadly. 

\subsection{Developing and using emulators}
There are numerous claims in the literature that the model-agnostic nature of emulators enables them to be trained on a wide range of different problems. We assert that in practice, this is almost never the case\footnote{Perhaps outside of Foundation Models, which are not the subject of this article}: practical applications of emulators require a keen understanding of the idiosyncrasies of the original model, and designing an emulator that can successfully tackle these model-specific behaviours will almost certainly be not universal. As an example, \citet{2025ApJ...991..169O} is able to achieve median performance of $1\%$ (but deviating up to 10\%) for light-curves, but also feature considerably larger model systematics (shown by fairly wide posteriors) due to their use-case of producing relatively smooth light-curves of pulsars. Small discrepancies of up to $10\%$ or more thus don't have a drastic effect on inferred posteriors. This contrasts BR25 in which the very tight posteriors that the underlying model can infer is then systematically more affected by larger errors. Indeed, in BR25, we suggest that the approach we took should be translatable to other models, but found that this was not true, even for a portion of the same model! It is important to ensure that the model to be approximated is similar enough in nature to that of the literature's, as an analogous problem will translate much better.

Another factor to consider is the breadth of the parameter space to be emulated. Cosmology has successfully deployed emulators for a huge range of problems with the caveat that many of the models emulated have very tight parameter constraints from previous work and methodologies, allowing for the range of model outputs to be well-constrained and relatively less varied. This does not translate as well to emulation with the goal of inferring from spectra from a variety of targets with a wide breadth of possible physical parameter ranges. This can mean strategies employed in this subfield do not translate as well (though we always recommend investigating if this is true as there are often a wider variety of ideas in the cosmological context due to a more mature adoption of machine learning techniques).

In particular, one should also take into consideration the type of features present in the model to be emulated. As discussed in this paper, the variety of high and low frequency features present in spectral models like \texttt{reltrans} are not conducive to convolutional architectures, and require Fourier embedding of the parameters to meaningfully reproduce high frequency features. 

We reiterate that no neural network seems to ever be universally able to obtain sufficient performance for all functions, and that bespoke systems should always be considered for each problem rather than attempting to find a universal solution. Researchers should rigorously investigate the model they wish to emulate, identify symmetries, correlations that can be analytically derived in a computationally efficient way, and the type of features in the model, and only then, if deemed worthwhile, begin to approach emulation. This approach also should help focus searching for analogous problems in the literature to derive inspiration from.

\subsection{A modular approach to emulation and future work}

In this paper, instead of emulating the entire model as done in BR25, here we took the the approach of only emulating part of the model. This yielded overall higher performance, because any imprecision in the emulator only contributes a small imprecision to the total spectrum. Here, the modular approach made the problem more complex to emulate, as a significant part of the non-linear behaviour was previously hidden by other parts of the model, which in turn means there is a trade-off between the complexity of the model component to be emulated compared to the gains in precision, but this is often not the case in many other complex models. 

In general, we suggest that this is a fruitful calculation to make: when reaching significant computational gains using end-to-end emulators is not possible without an overwhelming investment of resources, one can attempt to break down the model into computationally expensive and inexpensive constituents, and emulate only the computationally expensive parts, which may be significantly lower dimensional and easier to emulate. This makes them prime targets for approximation/emulation to achieve significant gains in computation time while reducing the effects of imprecision in emulation by retaining as much of the original model as possible.

In our example of \texttt{reltrans}, computation time is primarily driven by the ray-tracing required to calculate the relativistic kernel with the reflection spectrum from each portion of the disk. While in this paper, we emulate the total reflection spectrum after convolution, we could have taken a different approach: emulating the relativistic kernel itself.

To reduce the computation time required to convolve the relativistic kernel, reltrans performs the convolution of the reflection spectrum with the relativistic kernel in Fourier space (as the operation becomes simple multiplication). This does however then require a fast Fourier transform \citep{FFTW05,delsuc2024fourier} for each relativistic kernel and reflection spectrum. The relativistic kernel itself also takes a not insignificant time to compute as it requires numerical integration of light rays along the geodesics of the black hole. Once the kernel and Fourier transformed reflection spectra are multiplied, they must then be inverse Fourier transformed to obtain the expected observed spectrum. The number of FFTs performed thus becomes directly proportional to the resolution of the disk considered. In our case, the convolution between the kernel and rest-frame spectrum also takes approximately half the model's compute time and is therefore not as worth it to only replace the kernel instead of the entire end-to end reflection spectrum.

The new operator learning approach taken within this work is also a significant advancement towards emulation of timing products: timing products not only require an energy dependency but also a frequency dependency which becomes near impossible to emulate in a scientifically useful manner when using fixed grids in both energy and frequency. In the current version of \texttt{reltrans}, the lamppost corona approximation the model assumes is quite fast, but future models incorporating extended coronal geometries will become considerably slower without the parameter space increasing considerably in complexity. As \texttt{reltrans} and \texttt{RTFAST2} are updated in tandem, we expect for this emulator approach to become even more valuable.

\section{Conclusions}

We have explored emulation of the highly non-linear numerical model \texttt{reltrans} and used it to discuss the use case of end-to-end emulation for 1 dimensional spectral modelling. We show that understanding model behaviour with respect to individual parameters is an important diagnostic for understanding how challenging it may be to emulate a function with machine learning techniques. We develop a state-of-the-art emulator for the relativistically convolved reflection spectrum that is able to reproduce the spectrum to $\mathcal{O}(1)\%$ precision with $4-10$ speed-up in the most conservative conditions, and scaling considerably better in vectorized multiple evaluation cases. This emulator uses Fourier embedding to better model high-frequency and low-frequency features in the original model, as well as adopting a state-space approach, allowing the emulator to perform interpolation through the trained energy space for free. This modular approach allows emulation within \texttt{reltrans} to achieve better precision while still retaining much of the computational gains and has minimal effect on inferred model parameters, a direct improvement over previous work (BR25). This change has also enabled the emulator to be used across a larger energy bandpass, ranging from 0.1-100keV, practically making the emulator usable across all energies in which its assumptions are valid. \texttt{RTFAST2} will be directly implemented into \texttt{reltransDCp} to no longer require a bespoke python interface. This will mean that \texttt{RTFAST2} can be used in any software that is compatible with \texttt{xspec} formatted models.

We believe that the modular approach to emulation used in this paper is a promising strategy more generally, and encourage readers who wish to develop their own emulators to investigate their models systematically to identify similar bottlenecks similar to that of our own case of \texttt{reltrans}. 

\section*{Acknowledgements}

The authors would like to thank the BlackSTAR collaboration for useful discussion regarding \texttt{reltrans}. This work used the Dutch national e-infrastructure with the support of the SURF Cooperative using grant no. EINF-9197 and EINF-13764. G.M. acknowledges financial support from the European Union’s Horizon Europe research and innovation program under the Marie Sk\l{}odowska-Curie grant agreement No. 101107057 and the INAF Grant BLOSSOM. AI is supported by the European Union (ERC, X-MAPS, 101169908). Views and opinions expressed are however those of the author(s) only and do not necessarily reflect those of the European Union or the European Research Council. Neither the European Union nor the granting authority can be held responsible for them.

\section*{Data Availability}

All data was generated with \texttt{reltrans} v2.3.1, which is publically available at \url{https://github.com/reltrans/reltrans}. The data and parameters used in this paper can be found at \url{https://doi.org/10.5281/zenodo.21167981}. The use of \texttt{reltrans} was facilitated by \texttt{nDspec}, which is publically available at \url{https://github.com/nDspec/nDspec}. The python implementation of \texttt{RTFAST2} can be found at \url{https://github.com/API-AXIOM/RTFAST}. We anticipate merging \texttt{RTFAST2} into \texttt{reltrans} directly in the future. The code to make the plots in this paper can be found at \url{https://github.com/API-AXIOM/RTFAST2.0-paper-figures}.

\section*{Conflict of Interest}

Authors declare no conflict of interest.



\bibliographystyle{rasti}
\bibliography{biblio}




\appendix

\section{\texttt{reltrans} changes} \label{app:reltrans}

\begin{figure}
    \centering
    \includegraphics[width=\linewidth]{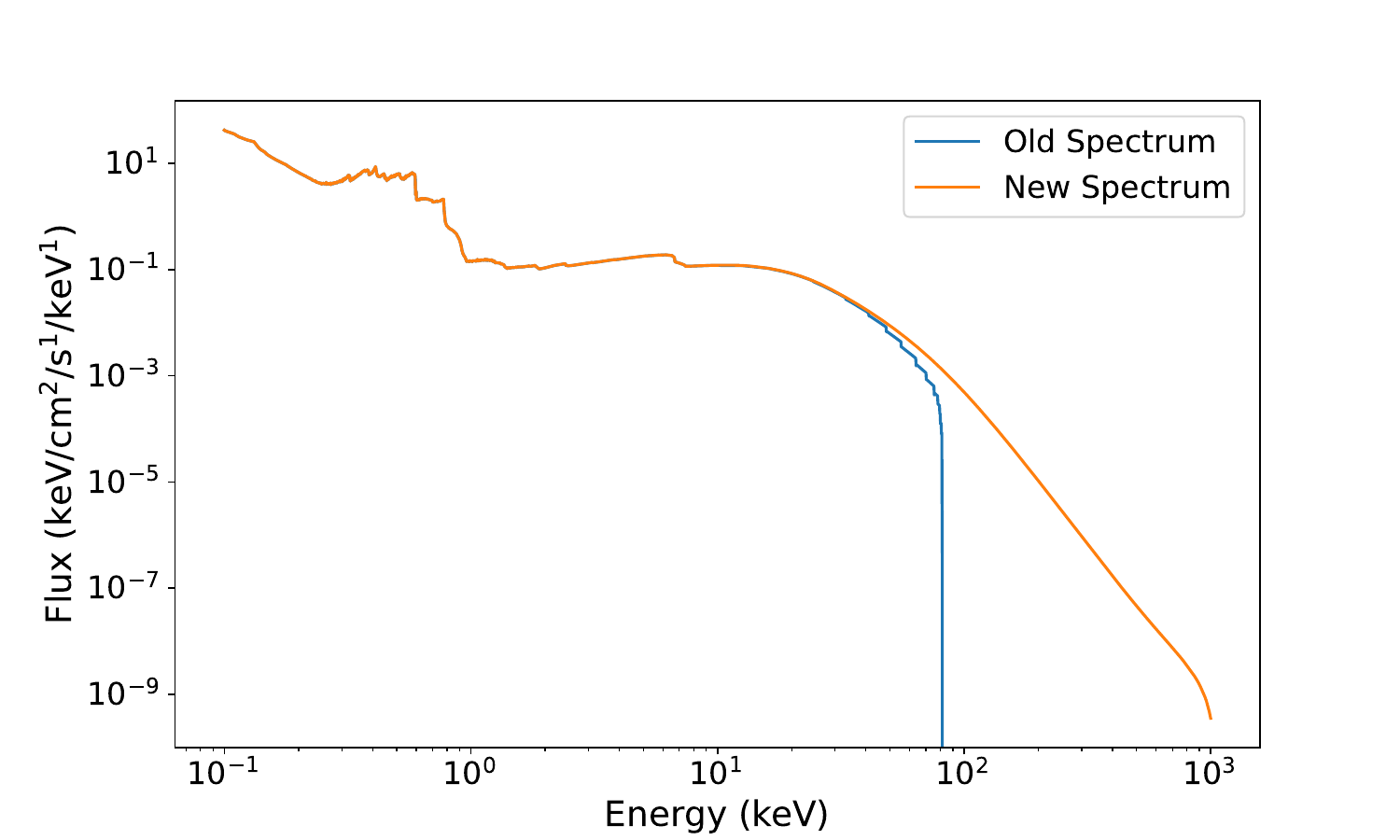}
    \caption{Comparison of the old reflection spectrum in blue vs the new reflection spectrum after changing the relativistic convolution method.}
    \label{fig:convolve}
\end{figure}

During generation of data for the emulator, we identified a bug in \texttt{reltrans} that meant that the reflection spectrum calculated by reltrans would sharply drop to 0 at high energies, with the drop-off dependent on $\Gamma$. This was identified to be caused by red noise from the Fourier transformation of the kernel and emitted reflection spectrum causing relatively small values of flux (in respect to the brightest bin) to be overwhelmed by noise. We explain the changes made to reduce this problem in the following section.

When we calculate the transfer function, we evaluate it by convolving the rest frame reflection spectrum with a kernel. In the simplest example of calculating the time-averaged spectrum, the observed specific flux is
\begin{equation}
F(E) = \int_\Omega \mathcal{R}(E/g) ~\epsilon~g^3~d\Omega.
\label{eqn:FEint}
\end{equation}
Here, $F(E)$ is a specific \textit{energy} flux, i.e. it is in units of photons per cm$^2$ per second, $\mathcal{R}$ is the emergent specific intensity (i.e. the rest frame reflection spectrum in units of photons per cm$^2$ per second per steradians), $\epsilon(r)$ is emissivity profile, $g=E/E_{\rm em}$ is the blueshift, and $d\Omega$ is the solid angle subtended by each patch of the disk. This can be expressed as a convolution
\begin{equation}
    F(\log E) = \mathcal{R}(\log E) \otimes W_\delta(\log E),
\end{equation}
where the kernel is
\begin{equation}
    W_\delta(E) = \int_\Omega \delta(E-g) ~\epsilon~g^3~d\Omega.
\end{equation}
We compute this convolution by Fourier transforming $\mathcal{R}(\log E)$ and $W_\delta(\log E)$, multiplying the two together, then inverse Fourier transforming the resulting product. This saves a lot of computational expense: defining $N$ as the number of bins in the energy grid, the cost of a manual convolution goes as $N^2$ whereas the cost of the Fourier transform convolution goes as $N \ln N$. The problem is that noise from the Fourier transform method can become dominant under certain conditions. This is particularly a problem when the rest-frame reflection spectrum is steeper than $E^{-2}$, in which case red noise leak becomes significant. 

The solution is to multiply the rest-frame reflection spectrum by a power law to stop it from being so steep. Multiplying both sides of Equation \ref{eqn:FEint} by $E^{\Gamma-1}$ gives
\begin{eqnarray}
    E^{\Gamma-1} F(E) &=& \int_\Omega E^{\Gamma-1}~\mathcal{R}(E/g) ~\epsilon~g^3~d\Omega \nonumber \\
    E^{\Gamma-1} F(E) &=& \int_\Omega (E/g)^{\Gamma-1}~\mathcal{R}(E/g) ~\epsilon~g^{2+\Gamma}~d\Omega \nonumber \\
    E^{\Gamma-1} F(E) &=& \int_\Omega \mathcal{H}(E/g) ~\epsilon~g^{2+\Gamma}~d\Omega,
\end{eqnarray}
where $\mathcal{H}(E) \equiv E^{\Gamma-1} \mathcal{R}(E)$. We can therefore evaluate $F(E)$ as $F(E) = E^{1-\Gamma} H(E)$, where
\begin{equation}
    H(\log E) = \mathcal{H}(\log E) \otimes \tilde{W}_\delta(\log E),
\end{equation}
and
\begin{equation}
    \tilde{W}_\delta(E) = \int_\Omega \delta(E-g) ~\epsilon~g^{2+\Gamma}~d\Omega.
\end{equation}

We now compute the FT of $\mathcal{H}(E)$ and $\tilde{W}_\delta(E)$, instead of $\mathcal{R}(\log E)$ and $W_\delta(\log E)$. $\mathcal{H}(E)$ is roughly constant with energy, whereas $\mathcal{R}(E)$ is roughly $\propto E^{1-\Gamma}$. Thus $\mathcal{H}(E)$ does not suffer as much from red noise leakage. Doing it this way does mean that the kernel is a steeper function of energy, but this matters considerably less due to it being a reasonably narrow function. 

The resulting improvement is plotted in Fig. \ref{fig:convolve} where the sudden anomalous drop (and steps in flux) in flux on the old blue spectrum instead continues on as expected all the way to 1000keV in the new orange spectrum.

\section{PCA Representations} \label{sec:representation}

\begin{figure*}
    \centering
    \includegraphics[width=0.95\linewidth]{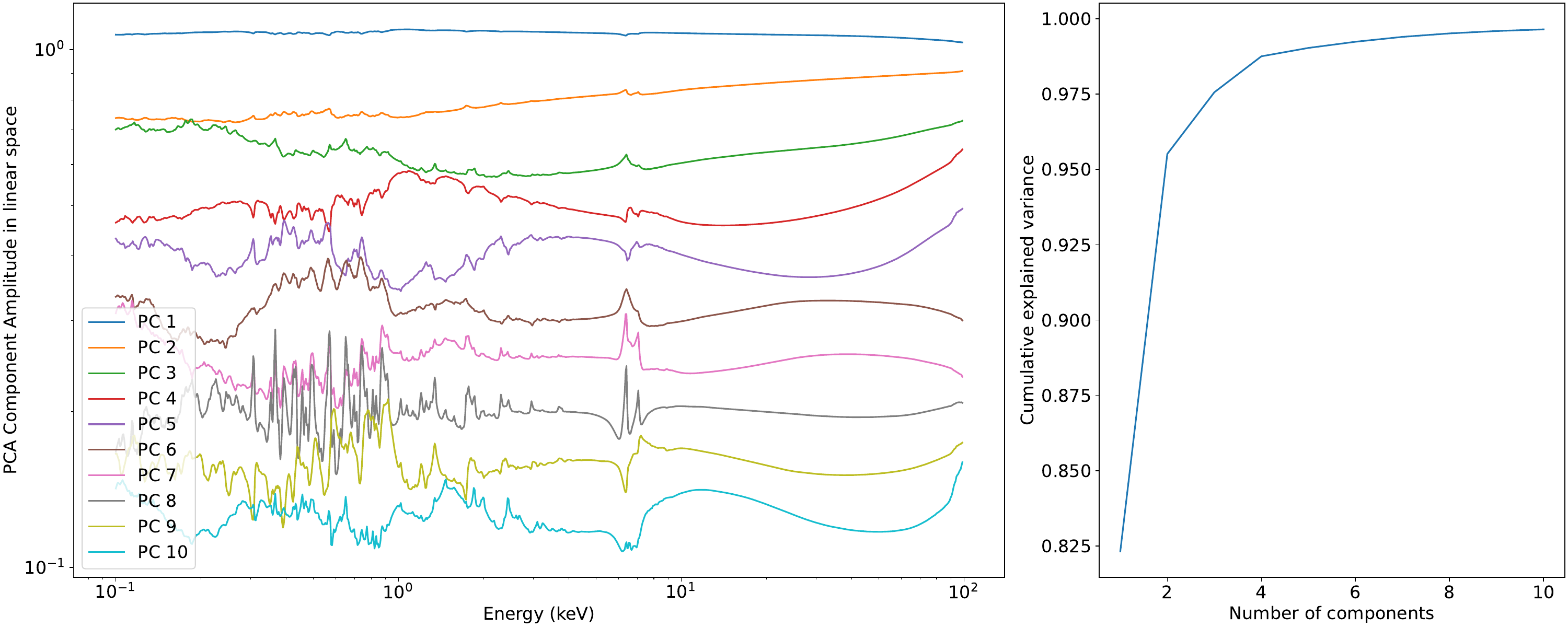}
    \caption{Left: Each line plots a PCA component from the trained PCA decomposition. Each is offset by a factor for visualisation purposes, with the first component at the top, the second component below the first and so forth. Right: cumulative explained variance of PCA decomposition of reltransDCp generated data as a function of the number of PCA components.}
    \label{fig:pca}
\end{figure*}

In BR25, we utilised Principal Component analysis (PCA) to compress the spectrum into a lower-dimensional representation and improve performance. This improved stability in training considerably, and using vector representations of the original data meant that correlations across the spectrum were preserved and encoded into the emulator very cheaply. It also serves as a useful way to understand variance in the model, as PCA decomposes the original data into vectors of most to least variance. However, we found that performing PCA decomposition did not help when emulating only the reflection spectrum. Here, we discuss how PCA decomposition for the reflection spectrum can be compared to that of BR25, why this was less effective and how PCA can be useful in identifying features in model data that can be used to inform emulation strategy.

In Fig \ref{fig:pca}, we show the first 10 component vectors of a PCA decomposition of our dataset in the left panel and the cumulative explained variance ratio in the right panel. Similar to that in BR25, the first component generally encodes an overall normalising factor. The second component mostly encodes a power-law like component as well as high-frequency components\footnote{Here, we use ``high-frequency'' to denote rapid variations of the spectrum with energy, for example caused by narrow emission and absorption lines, as opposed to low-frequency features such as overall trends or broad spectral features} at low energies (associated with the so-called "soft excess" of emission and absorption lines) as well as a relativistically smeared iron-like feature at 6.4keV. Additional components represent various features of the complex lines as well as the relativistic wings of the Fe line.  While not shown here, to recreate the spectrum to sufficient precision ($\leq 0.1\%$), only around $10^{-6}$ of the variance could be left unexplained which requires 200 PCA components (1/5 of the total dimensions in the data). Later components included oscillatory features in order to recreate what should be smooth, non-linear functions (such as in the Compton hump).

\begin{figure}
    \centering
    \includegraphics[width=0.98\linewidth]{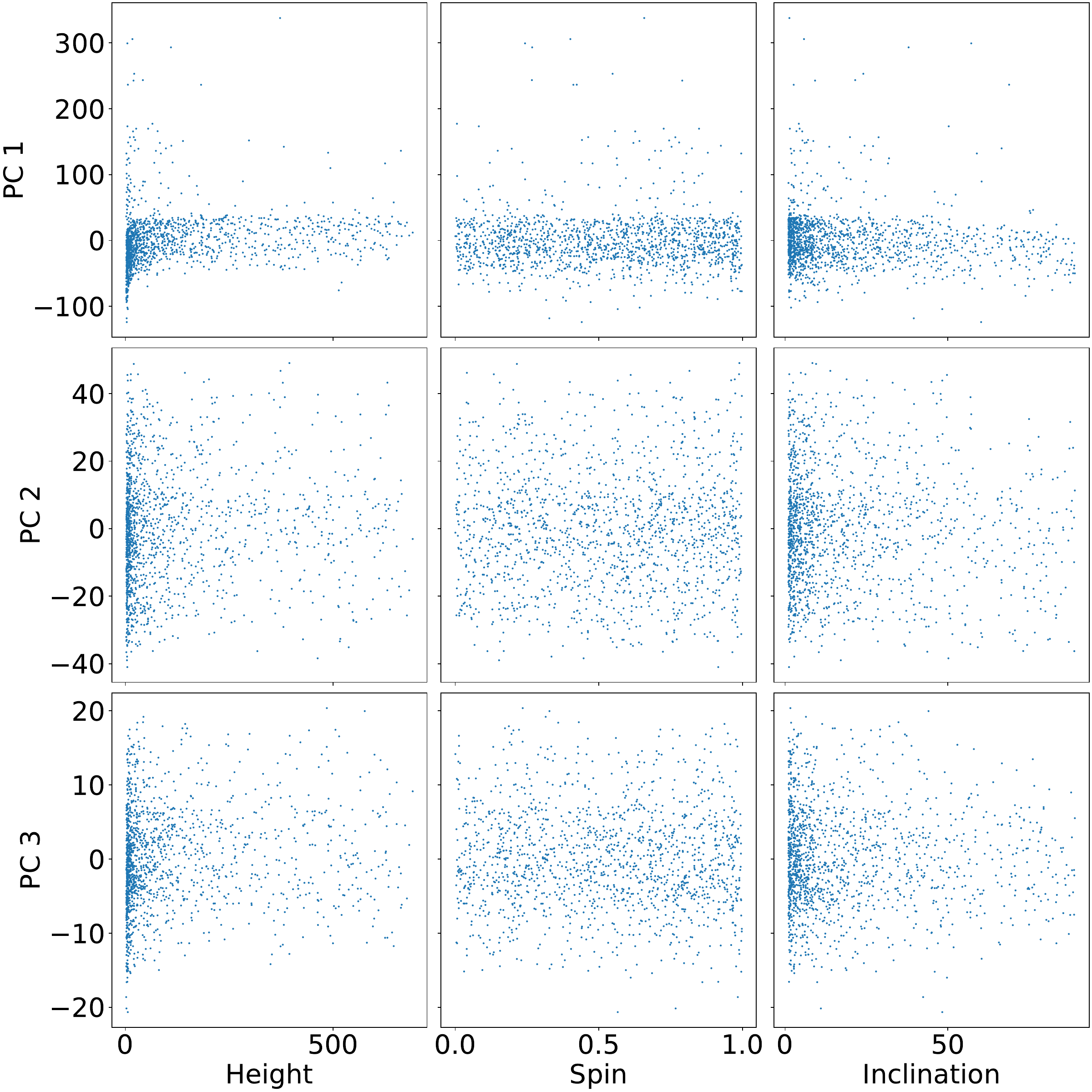}
    \caption{Scatter plots of PCA vector magnitude vs height, spin and inclination in the first, second and third column respectively.}
    \label{fig:pca_scatter}
\end{figure}

In Fig \ref{fig:pca_scatter}, we show the first 3 PCA vectors' magnitude correlated with 3 parameters (height, spin and inclination). There is almost no correlation in 1-D space between these parameters and vectors. This likely means that the relationship between these vectors and parameters (if such a relationship exists) tracks a line through multi-dimensional parameter space. In these cases, emulation becomes more difficult, either requiring more data or either a more complex or larger network architecture. In cases where almost as many PCA components are required as the dimensionality of the data to reproduce the model outputs to sufficient precision, it is often better to simply approximate the original model output directly. If computation time is a significant constraint, data often must become the driver of improvement in the model. 

Analyses such as discussed above can be helpful in informing what type of emulation strategy may work for your use case. In general, if the function to be approximated is very smooth and has clear non-degenerate changes in the function outputs, a very simple neural network (even as extreme as a single layer feed-forward neural network) will likely be able to represent the function in question, as long as enough data is used during training. In these cases, representations like PCA are often not even required. 

We also tried other strategies of representing the spectrum: negative matrix factorisation, wavelets and auto-encoders. We found negative matrix factorisation to perform significantly worse in recreating the spectrum, introducing artifacts into the reconstruction and being less accurate. This swiftly eliminated it as a viable representation strategy. Wavelets were able to perfectly reconstruct the spectrum, but yielded no actual gains in reducing the size of the output features, and were not more interpretable for neural networks either. Finally, auto-encoders yielded poor performance in recreating the spectra from itself. This makes logical sense in this context. Auto-encoders attempt to find a lower-dimensional latent space that is representative of the data. Indeed, they are sometimes used in an unsupervised manner to attempt to find parameters that can be used to directly model data. However, in this case, we already have the perfect latent space, the original parameters used to generate the model. If an auto-encoder of sufficient precision was possible to train, we also would not have trouble with creating an emulator of sufficient precision. It should also be noted that by choosing to emulate the reflection spectrum, instead of the total spectrum, we have already exploited the most obvious symmetries within the model.

Having shown that linear representations fail to exploit structure in the model, we may now ask what a bespoke architecture must do instead to solve the problem of building an emulator for this type of model.

\section{Chi squared analysis} \label{sec:chi2}

\begin{figure}
    \centering
    \includegraphics[width=\linewidth]{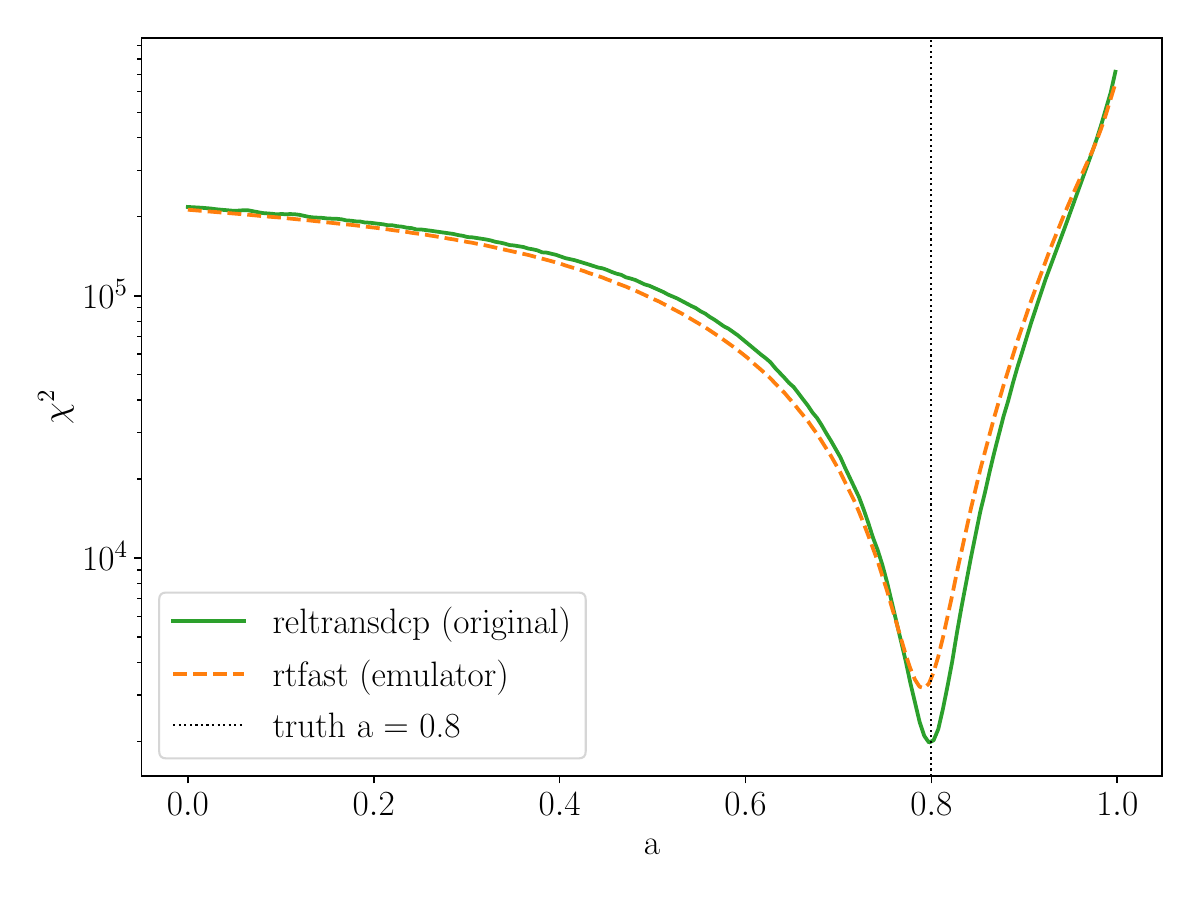}
    \caption{Plot of chi-squared statistic as a function of the parameter of spin when all other parameters are fixed. The solid green line denotes \texttt{reltrans} (the original numerical model) and the dashed orange line denotes \texttt{RTFAST}. The true parameter of the simulation is plotted as a vertical dotted black line.}
    \label{fig:chi2spin}
\end{figure}

\begin{figure}
    \centering
    \includegraphics[width=\linewidth]{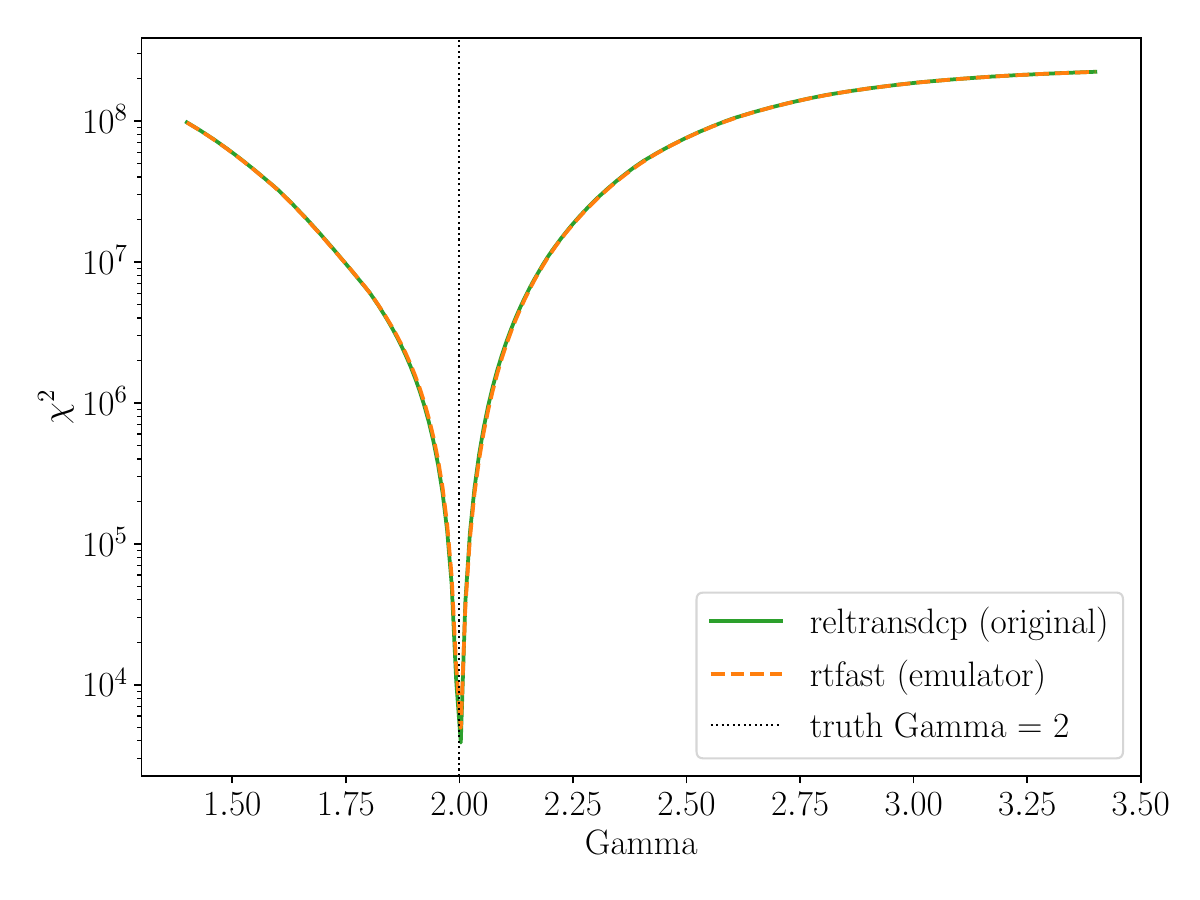}
    \caption{Plot of chi-squared statistic as a function of the parameter of $\Gamma$ when all other parameters are fixed. The solid green line denotes \texttt{reltrans} (the original numerical model) and the dashed orange line denotes \texttt{RTFAST}. The true parameter of the simulation is plotted as a vertical dotted black line.}
    \label{fig:chi2gamma}
\end{figure}

In Figures \ref{fig:chi2spin} and \ref{fig:chi2gamma}, we show the chi-squared statistic as a function of spin and $\Gamma$ respectively for both \texttt{reltrans} and \texttt{RTFAST} when compared against the simulated data in Section \ref{sec:MCMC}. All other parameters are fixed to their true parameters. We see that in both plots that the minimum chi-squared is recovered in the same place. While the chi-squared appears almost entirely identical in the case of $\Gamma$, the minimum chi-squared seems much higher in the case of spin, but this is due to the small changes in chi-squared when varying spin compared to the larger changes in $\Gamma$ making apparent differences much larger to the human eye. Importantly, we observe that the minimum point in parameter space is nearly identical, indicating low to minimal bias imposed by the model, particularly compared to BR25.

The shallower likelihood space present in RTFAST may explain the faster convergence, however systematic enough analysis to understand \texttt{reltrans}'s likelihood space has yet to be performed to make a meaningful conclusion on the difference.

\section{MCMC diagnostics} \label{sec:MCMC}

\begin{figure}
    \centering
    \includegraphics[width=\linewidth]{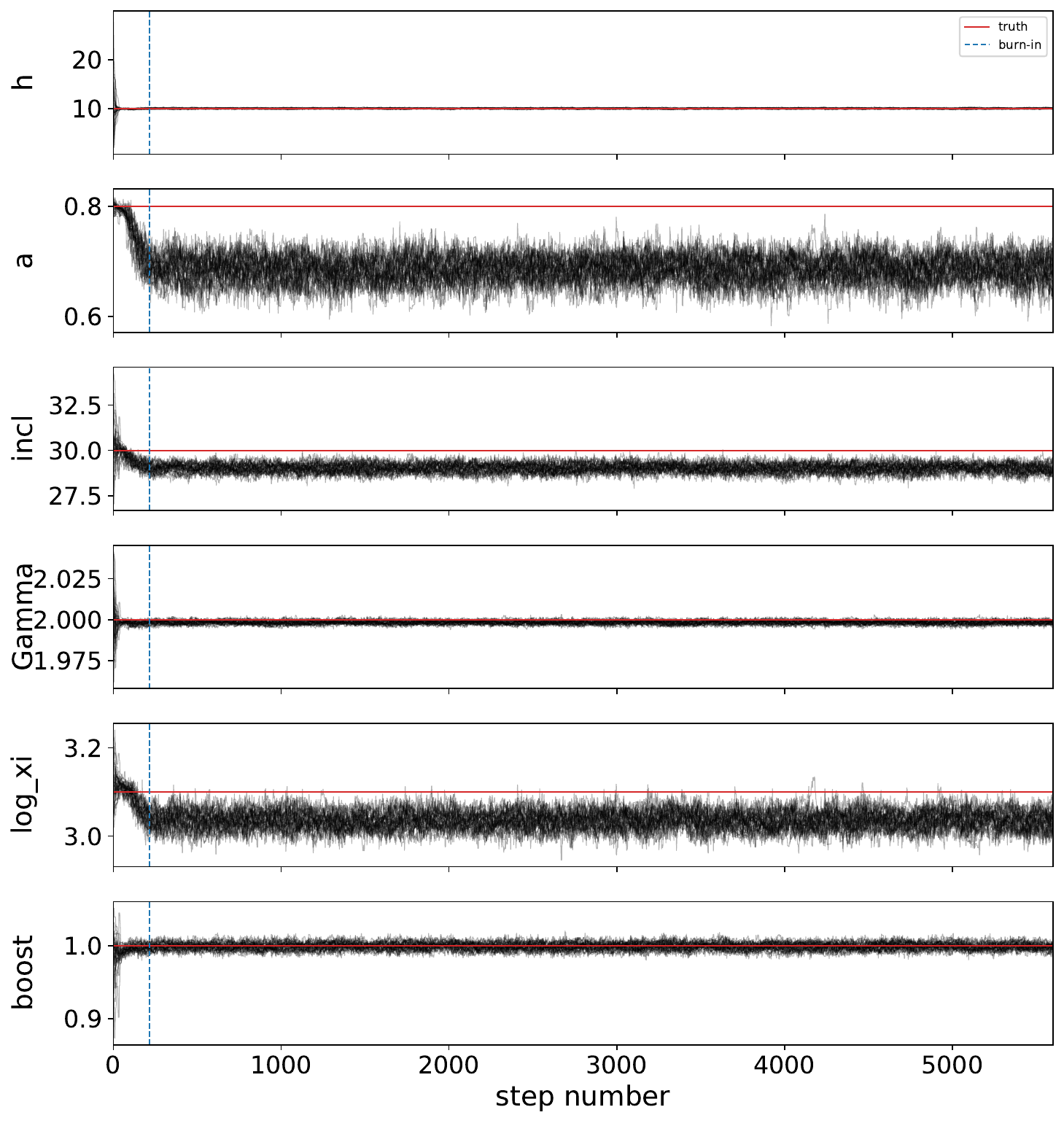}
    \caption{Trace plots of the MCMC walkers. The truth for each parameter is plotted as a horizontal red line.}
    \label{fig:trace}
\end{figure}

\begin{figure}
    \centering
    \includegraphics[width=\linewidth]{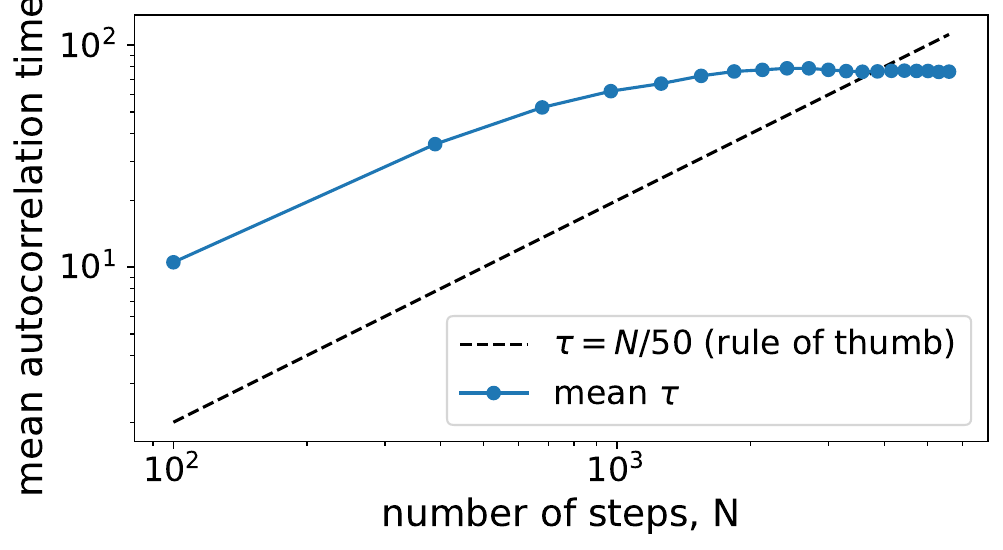}
    \caption{Auto-correlation length of the MCMC ensemble as a function of steps. The diagonal dashed line plots the 50 times auto-correlation length rule of thumb.}
    \label{fig:auto-corr}
\end{figure}

In Fig. \ref{fig:trace}, we plot the trace plots of the MCMC performed in Section \ref{sec:ppc}. 2 of the walkers in this MCMC were thrown away when generating the posterior plots due to getting stuck in a local minimum that was approximately 500 times less probable than the main mode. This issue is somewhat common for MCMC inference with \texttt{reltrans} as well.

We also plot the auto-correlation length as a function of steps in Fig. \ref{fig:auto-corr}. We reach reliable estimates of the auto-correlation length at approximately 5600 steps due to reaching 50 times the estimate as recommended by \citet{foreman2013emcee}.


\changes{\section{Distribution of emulator errors across parameter space} \label{app:failures}}

\begin{figure*}
    \centering
    \includegraphics[width=\linewidth]{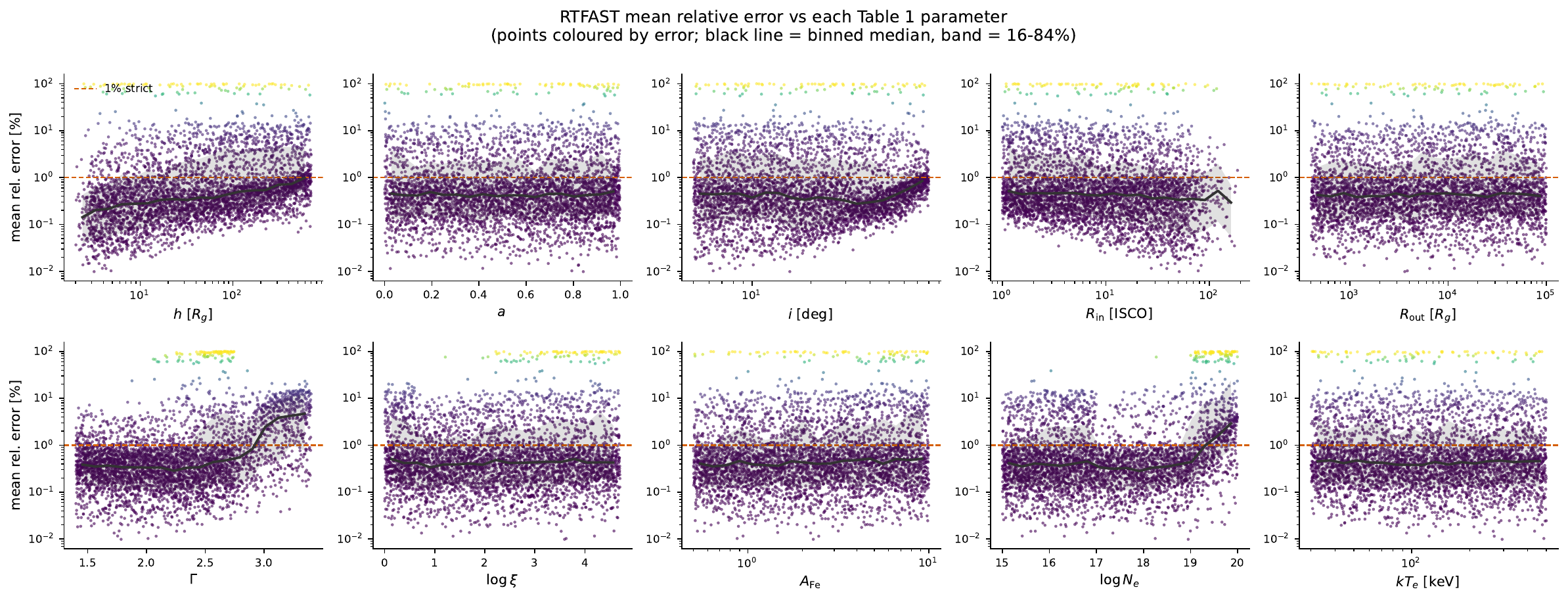}
    \caption{\changes{Mean relative error per spectrum against each model parameter. Points are coloured by the magnitude of the error.}}
    \label{fig:hcscan_param}
\end{figure*}

\begin{figure*}
    \centering
    \includegraphics[width=\linewidth]{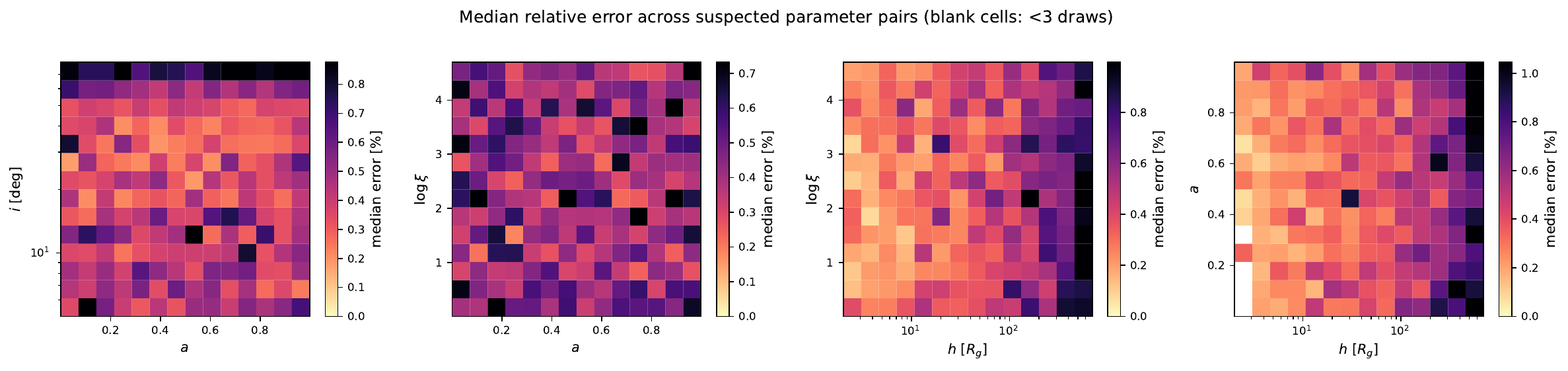}
    \caption{\changes{Mean relative error across pairs of parameters. Edge-on and very high coronae trend to have the worst reconstruction.}}
    \label{fig:hcscan_heat}
\end{figure*}

\begin{table}
\centering
\begin{tabular}{lrrr}
\hline
Region & $<1\%$ & $<10\%$ & $N$ \\
\hline
All draws        & 75.0\% & 95.2\% & 4095 \\
$a > 0.98$       & 66.7\% & 95.7\% & 93 \\
$i > 75^{\circ}$ & 63.2\% & 97.9\% & 95 \\
$\log \xi > 4$   & 74.2\% & 91.4\% & 590 \\
$h < 3 R_g$      & 90.0\% & 96.0\% & 100 \\
$\Gamma < 1.6$   & 86.2\% & 100.0\% & 492 \\
$\Gamma > 3.0$   & 22.9\% & 83.8\% & 468 \\
\hline
\end{tabular}
\caption{\changes{Fraction of test spectra meeting the $1\%$ and $10\%$ mean relative error criteria, by region of parameter space, from a Sobol scan of 4095 draws over the training domain.}}
\label{tab:failures}
\end{table}

\changes{The errors quoted in Section \ref{sec:sota} are fractions of the energy space rather than of test spectra. To characterise performance per spectrum, and to establish where in parameter space the emulator is least accurate, we generated an additional test set over the full training domain using a Sobol sequence of 4095 draws and compared \texttt{RTFAST2} against \texttt{reltransDCp} for each. We characterise each spectrum by its mean relative error across the analysis band (1.62-99.98 keV; grid bins outside the 0.1-100 keV training range are excluded). The median spectrum has a mean error of $0.43\%$ and the 90th percentile is $5.0\%$, with $75.0\%$ of spectra falling below the $1\%$ criterion and $95.2\%$ below $10\%$.}

\changes{In Fig. \ref{fig:hcscan_param}, we plot the mean error of each spectrum against each parameter, and in Fig. \ref{fig:hcscan_heat} across pairs of parameters. Spearman rank correlations between the mean error and each parameter give $\rho=+0.387$ for $h$ and $\rho=+0.316$ for $\Gamma$, while spin is effectively uncorrelated ($\rho=+0.009$), as are inclination ($+0.064$) and ionisation ($+0.022$). The fraction of spectra meeting each criterion by region of parameter space is listed in Table \ref{tab:failures}. The dominant region of poor performance is soft spectra, $\Gamma>3$, where only $22.9\%$ of draws meet the $1\%$ criterion. Very high spin and edge-on inclination show modest degradation ($66.7\%$ and $63.2\%$ respectively), but with 93 and 95 draws these differences are only $1.7\sigma$ and $2.4\sigma$ from the global rate, and no monotonic trend with either parameter is present across the full range. Extreme ionisation is indistinguishable from the global rate, while small values of $h$ are better than average at $90.0\%$.}

\changes{The common feature of the failing parameter sets is low reflected flux. Soft $\Gamma$ and large $h$ both suppress the reflection spectrum, so the relative error grows in regions that contribute little to the total counts. Weighting the error of each spectrum by its own flux supports this: the flux-weighted 90th percentile is $2.4\%$ against $5.0\%$ unweighted, while the medians are nearly identical ($0.459\%$ against $0.430\%$). The largest relative errors therefore occur where the model contributes least to the likelihood.}

\changes{Finally, we should note that the filters on parameter space imposed in Section \ref{sec:trainingdata} show in Fig. \ref{fig:hcscan_param} in the sharp increase in error rate at $\log{N_e}>19$. In a future iteration of the emulator, we will likely remove the filter imposed to prevent the mirror-like Compton reflecting spectra from \texttt{xillver} contaminating the data set and instead self-consistently account for it in our emulator and training data design.}

\bsp	
\label{lastpage}
\end{document}